\pdfoutput=1
\documentclass[aps,prd,12pt,nofootinbib]{revtex4}
\usepackage{epsfig}
\usepackage{graphicx}
\usepackage[font=small,skip=0pt,justification=raggedright]{caption}
\usepackage{amsmath}
\usepackage{amssymb}
\usepackage{mathrsfs}
\usepackage{verbatim}
\usepackage{dutchcal}
\usepackage{comment}

\usepackage[normalem]{ulem}
\usepackage{xcolor}

\newcounter{fig}   \newcommand{\lbfig}[1]{\refstepcounter{fig}
\label{#1} }

\newcommand{\bea}{\begin{eqnarray}}
\newcommand{\eea}{\end{eqnarray}}
\newcommand{\be}{\begin{equation}}
\newcommand{\ee}{\end{equation}}

\def\pa{\partial}

\def\({\left(}
\def\){\right)}

\newcommand{\re}[1]{(\ref{#1})}

\newcommand{\tr}{\mbox{Tr}}

\def\rlx{\relax\leavevmode}
\def\IR{\rlx\hbox{\rm I\kern-.18em R}}
\def\one{\hbox{{1}\kern-.25em\hbox{l}}}

\newcommand{\eqn}{\begin{eqnarray}}
\newcommand{\eqnx}{\end{eqnarray}}

\date{\today}

\begin{document}
\title{
Skyrmions and pion stars\\ in the $U(1)$ gauged Einstein-Skyrme model}

\author{R.~Kirichenkov}
\affiliation{Belarusian State University, Minsk 220004, Belarus}
\author{J.~Kunz}
\affiliation{Institute of Physics,
Carl von Ossietzky University Oldenburg, Germany
Oldenburg D-26111, Germany}
\author{Nobuyuki Sawado}
\affiliation{Department of Physics and Astronomy, Tokyo University of Science, Noda, Chiba 278-8510, Japan}
\author{Ya.~Shnir}
\affiliation{BLTP, JINR, Dubna 141980, Moscow Region, Russia\\
Hanse-Wissenschaftskolleg, Lehmkuhlenbusch 4, 27733 Delmenhorst, Germany\\
Institute of Physics,
Carl von Ossietzky University Oldenburg, Germany
Oldenburg D-26111, Germany}

\begin{abstract}
We consider topological and non-topological regular soliton solutions in the Einstein-Maxwell-Skyrme theory.
We analyze the properties of these solutions and determine their domains of existence.
The dependence of the solutions on the gauge coupling and on the strength of the effective gravitational coupling are examined.
Topologically trivial localized field configurations, \textit{pion stars}, are shown to exist, as non-linear gravitational bound states of the Skyrme field.
Both spherically-symmetric and axially-symmetric pion stars are considered.
We find that these solutions share many features with the usual (mini-)boson stars.
In particular they also exhibit a spiraling behavior and do not possess a flat space limit.

\end{abstract}
\maketitle

\section{Introduction}
The Skyrme model \cite{Skyrme:1961vq,Skyrme:1962vh} is a modified version of the nonlinear sigma model in $3+1$ dimensional spacetime.
It serves as a simple prototype of a theory supporting topological solitons (for reviews, see, for example \cite{Manton:2004tk,Manton:2022,Brown:2010api,Shnir:2018yzp}). Originally it was conjectured that baryons can be considered as topological solitons.
Thus the baryon number was identified with the topological degree of the field configuration.
In this approach pions correspond to the linearized fluctuations of the Skyrme field.
A quasi-classical quantization of the rotational and isorotational degrees of freedom of Skyrmion solutions leads to predictions of the physical properties of nuclei, that are in a reasonable agreement with experimental data \cite{Manton:2022,Brown:2010api}.

The Skyrme model has received much attention in a variety of fields.
One of these interesting developments is related to the study of self-gravitating Skyrmions \cite{Luckock:1986tr,Glendenning:1988qy,Heusler:1991xx,Bizon:1992gb,Heusler:1993ci}.
In such a context, the Einstein-Skyrme model can be considered as a model of compact stars, exhibiting a spiraling behavior as neutron stars or boson stars beyond the maximum mass, indicating their expected collapse to a black hole (see e.g., \cite{Harrison:1965}).
Moreover, the Einstein-Skyrme model provided an early counterexample to the no hair conjecture \cite{Luckock:1986tr,Droz:1991cx} (see also \cite{Kleihaus:1995vq,Shiiki:2005xn,Sawado:2004yq,Brihaye:2005an} and reviews \cite{Volkov:1998cc,Volkov:2016ehx}).

Apart from topological solitons, a distinct class of localized field configurations in flat space is given by non-topological solitons or Q-balls \cite{Rosen:1968mfz,Friedberg:1976me,Coleman:1985ki}.
Such solutions may exist in models possessing an unbroken global symmetry.
Typical examples are the so-called Friedberg-Lee-Sirlin two-component model with a symmetry breaking potential \cite{Friedberg:1976me} and the model with a single complex scalar field and a suitable self-interaction potential \cite{Coleman:1985ki}.

On the other hand, stable localized soliton-type configurations, so called {\emph{boson stars}}, may arise when the complex scalar field is coupled to gravity \cite{Kaup:1968zz,Feinblum:1968nwc,Ruffini:1969qy}.
Some of these boson stars are linked to the corresponding flat space non-topological solitons and Q-balls.
Solutions of another type, like the boson stars of Einstein-Klein-Gordon theory, do not possess a flat space limit, however.
Similar topologically trivial solutions with harmonic time dependence exist in the Einstein-Skyrme model \cite{Ioannidou:2006nn} and in the $O(3)$-sigma model coupled to gravity \cite{Verbin:2007fa,Herdeiro:2018daq,Cano:2023bpe}.

Several modifications of the Skyrme model have been proposed throughout the last two decades with the aim to improve phenomenological predictions of the theory.
In particular, in order to account for the explicit breaking of isospin symmetry, it has been suggested to consider the $U(1)$ gauged version of the Skyrme model \cite{Piette:1997ny,Radu:2005jp,Livramento:2023keg,Livramento:2023tmm}.
On the other hand, there are charged Q-balls and boson stars in extended Einstein-Maxwell-scalar theories with local $U(1)$ symmetry \cite{Jetzer:1989us,Jetzer:1992tog,Jetzer:1991jr,Pugliese:2013gsa,Kleihaus:2009kr,Kumar:2014kna,Kunz:2021mbm}.

We here investigate the properties of the $U(1)$ gauged regular self-gravitating solutions of the Einstein-Skyrme model, focusing our study on the Skyrmions of topological degree one, and on non-topological localised configurations, which we will refer to as the \textit{pion stars}, and determine their domains of existence.

This paper is organized as follows.
In Sec. II we introduce the model.
Here we discuss the gauge fixing, the parametrization of the metric and the matter fields, the physical quantities of interest and the boundary conditions under which the field equations are solved numerically.
In Sec. III we present the results of our study  of self-gravitating gauged Skyrmions of topological degree one, and the dependence of the solutions on the strength of the effective gravitational coupling constant.
Pion stars are discussed in Sec. IV.
We show that, besides axially symmetric pion stars there are also spherically symmetric pion stars.
The pion stars represent topologically trivial solutions which, similar to the (mini-)boson stars, do not possess a flat space limit.
We conclude with a discussion and final remarks in Sec. V.

\section{The model}

\subsection{Action}
We consider the $U(1)$ gauged Einstein-Skyrme model in $3+1$ dimensional spacetime defined by the action
\be
\label{LagSk}
S = \int d^4 x \sqrt{-\cal g}
\left(\frac{R}{16 \pi G} + {\cal L}_m \right)\ ,
\ee
where the gravity part is the usual Einstein-Hilbert action,
$\cal g$ is the determinant of the metric, $R$ is the curvature scalar and $G$ is Newton's constant.
The Lagrangian of the
matter fields ${\cal L}_m$ is given by the $U(1)$ gauged
$SU(2)$-Skyrme-Maxwell model \cite{Callan:1983nx,Piette:1997ny,Radu:2005jp,Livramento:2023keg,Livramento:2023tmm}

\be
\begin{split}
\label{Lag0}
{\cal L}_{m}&=- \frac{f_\pi^2}{16}\,\,\tr \(D_\mu U\,D^\mu U^\dagger\)+
\frac{1}{32\,a_0^2} \,\tr \(\left[D_\mu U\, U^\dagger,\, D_\nu U\,U^\dagger\right]^2\)+
\frac{m_\pi^2 \,f_\pi^2}{8}\,\tr\(U-\mathbb{I}\) -\frac{1}{4}\,{{\cal F}}_{\mu\nu}\,{{\cal F}}^{\mu\nu} \ ,
\end{split}
\ee
where $f_\pi$, $a_0$ and $m_\pi$ are parameters of the model with dimensions $[f_\pi] =L^{-1}$, $[a_0]= L^0$ and $[m_\pi] =L^{-1}$, respectively. The electromagnetic field-strength tensor is ${\cal F}_{\mu\nu} = \partial_\mu {\cal A}_\nu -\partial_\nu {\cal A}_\mu $, the covariant derivative of the $SU(2)$ valued Skyrme field $U$ is defined as
\be
D_\mu U = \pa_\mu U +i\,e\,{\cal A}_\mu\,\left[Q,\,U\right] \ ,
\ee
and the charge matrix is $Q\equiv\frac{1}{2}\(\frac{1}{3}\,\mathbb{I}+\tau_3\)={\rm diag}\,\(\frac{2}{3},\,-\frac{1}{3}\)$.

It is convenient to rescale the model by introducing the  dimensionless coordinate $x=(a_0 f_\pi/2) r$, the dimensionless mass parameter $m=2\,m_\pi/(a_0\,f_\pi)$ and the dimensionless gauge potential $A_\mu = (2/f_\pi) {\cal A}_\mu$.
The dimensionless field strength tensor $F_{\mu\nu}$ is then constructed from the gauge potential $A_\mu$ and the partial derivative with respect to the coordinate $x$, while the covariant derivative contains the scaled gauge coupling $g = e/a_0$.
The effective gravitational coupling constant\footnote{Note that, because of a different choice of the parameters of the Einstein-Skyrme model \re{LagSk}, our $\alpha^2$ differs by a factor of 2 from that defined in \cite{Ioannidou:2006nn}.} is $\alpha^2=\frac{1}{2}\pi G f_\pi^{2}$.

In terms of these units the Skyrme-Maxwell Lagrangian \re{Lag0} becomes
\be\begin{split}
\label{Lag0b}
{\cal L}_{m}&=-\frac{1}{2}\,\,\tr \(D_\mu U\,D^\mu U^\dagger\)+\frac{1}{16} \,\tr
\(\left[D_\mu U\, U^\dagger,\, D_\nu U\,U^\dagger\right]^2\)+m^2\,\tr\(U-\mathbb{I}\)
-\frac{1}{2}\,{{F}}_{\mu\nu}\,{{F}}^{\mu\nu}\ .
\end{split}
\ee

The requirement of finite energy leads to the restriction that the matrix-valued field $U$ approaches the vacuum at all points at spatial infinity, $U \xrightarrow[\vec r \to\infty]{} {\mathbb{I}}$, thus the Skyrme field becomes a map $U:S^3 \mapsto S^3$.
The corresponding topological current $B^\mu$ is
\be
B^\mu=\frac{1}{\sqrt{-\cal g}}\frac{1}{24\,\pi^2}\, \,\varepsilon^{\mu\nu\rho\sigma}\,\tr \(R_\nu\,R_\rho\,R_\sigma\) \, , \label{chargemain}
\ee
where $R_\mu=(\partial_\mu U)U^\dagger$ is the $SU(2)$-valued left-invariant current.
The corresponding charge $B=\int_\Sigma d^3 x  B^0 $ is interpreted as the baryon number.

The Skyrme field $U$ can be decomposed into the scalar component $\phi_0$ and the pion isotriplet $\phi_k$ via
\be
U=\phi_0\,\mathbb{I}+i\,\sum_{n=1}^3\phi_n\,\tau_n \ ,  \label{decomposition}
\ee
where $\tau_n$ are the usual Pauli matrices, and the field components
$\phi^a = (\phi_0,\phi_k)$ are subject to the sigma-model constraint,
$\phi^a \cdot \phi^a = 1 $.

In this component notation the Lagrangian for the $U(1)$ gauged Skyrme model \re{Lag0b} can be written as
\be
\begin{split}
\label{Lag}
{\cal L_m}=
 D_\mu \phi^a D^\mu \phi^a
&-\frac12 (D_\mu \phi^a D^\mu \phi^a)^2 +
\frac{1}{2} (D_\mu \phi^a D_\nu \phi^a)(D^\mu \phi^b D^\nu \phi^b)\\
&- 2\,m^2\,(1-\phi_0) - \frac{1}{2}\,F_{\mu\nu}\,F^{\mu\nu}\ ,
\end{split}
\ee
where
\be
D_\mu \phi^\alpha = \pa_\mu \phi^\alpha +g\,A_\mu\,\varepsilon_{\alpha\beta}\,\phi^\beta \,,
\qquad D_\mu \phi^A = \pa_\mu \phi^A, \qquad \alpha,\,\beta=1,\,2, \ \ A=0,\,3 \ .
\label{covariant}
\ee
The two components of the energy-momentum tensor are
\begin{eqnarray}
T^{\mu\nu}=T^{\mu\nu}_{(M)}+T^{\mu\nu}_{(S)}
\label{T}~,
\end{eqnarray}
where the electromagnetic contribution of the Maxwell term is
\begin{eqnarray}
T^{\mu\nu}_{(M)}=F^{\mu\sigma }\,F^{\nu}_{\:\:\sigma}-\frac{g^{\mu\nu}}{4}\,F_{\alpha\beta}\,F^{\alpha\beta} \ ,
\label{T_M}
\end{eqnarray}
and the stress-energy tensor of the gauged Skyrmion is
\begin{eqnarray}
\nonumber T^{\mu\nu}_{(S)}&=&2\,\left[ D^\mu \phi_a\,D^{\nu}\phi^a-\(D^{[\mu} \phi^a\,D^{\alpha]} \phi^b\)\,\(D^{[\nu} \phi_a D_{\alpha]}\phi_b\)\right]  \\
&&-g^{\mu\nu}\,\(\( D_\alpha \phi_a \)^2-\frac12 \( D_{[\alpha} \phi_a \,D_{\beta]} \phi_b \)^2- 2\,m^2\,(1-\phi_0)\) \ .
\label{stress}
\end{eqnarray}

\subsection{Gauge transformations}

The gauged Skyrme model \re{Lag} is invariant with respect to the local $U(1)$ gauge transformations
\be
U\rightarrow e^{-i\,g\,\frac{\xi}{2}\tau_3} \,U\,e^{i\,g\,\frac{\xi}{2}\tau_3}\,, \quad {\rm or}\quad
\phi_1+i\,\phi_2\rightarrow e^{i\,g\,\xi}\,\(\phi_1+i\,\phi_2\),
\qquad A_\mu \rightarrow A_\mu + \pa_\mu \xi \ ,
\label{gauge}
\ee
where $\xi$ is any real function of the coordinates.

The vacuum of \eqref{Lag} corresponds to $U=\one$, $D_\mu \phi^a  =0$ and $F_{\mu\nu}=0$.
In the stationary gauge, where no explicit time dependence of the fields is present, one can consider the vacuum boundary conditions \cite{Radu:2005jp}
\be
U\(\infty\)=\mathbb{I},\qquad   A_0(\infty) = V,\qquad\,A_i(\infty) = 0 \  ,
\label{asympcondition}
\ee
where $V$ is a real constant.
However, the asymptotic value of the electric potential $A_0(\infty)$ can be adjusted via the residual $U(1)$ degree of freedom.
In particular, the transformation \eqref{gauge} with $\xi = -V\, t$ allows us to set $A_0\(\infty\)=0$.
The components of the charged pion field then transform as $\phi_\alpha \rightarrow e^{-i\omega \,t}\,\phi_\alpha$, where $\omega=gV$, and thus the charged pion fields obtain an explicit time dependence with frequency $\omega$.
In other words, in the Skyrme-Maxwell model \re{Lag0} isorotations of the Skyrmion are associated with time-dependent gauge transformations \cite{Radu:2005jp}.
Note, that for both the $U(1)$ gauged and isospinning Skyrmions \cite{Battye:2005nx,Ioannidou:2006nn,Battye:2014qva}, the pion mass
term is necessary to stabilize the configurations.

The asymptotic expansion of the fields around the vacuum  \eqref{asympcondition} yields in the stationary gauge
\be
\approx \mathbb{I} + i\, \phi_k\,\tau_k +{\cal O}\,\(\phi_k^2\)\,,\qquad\qquad A_\mu=a_\mu +V\,\delta_{0\mu} +{\cal O}\,\(a_\mu^2\) \ ,
\label{flutuations}
\ee
and the linearized equations for the pion fields in the asymptotically flat region become
\be
\partial_r^2 \phi_{1,2} +
\frac{2\,\partial_r \phi_{1,2}}{r} -\(m^2-g^2 V^2\)\,\phi_{1,2}=0\ , \qquad\qquad
\partial_r^2 \phi_3 + \frac{2\,\partial_r \phi_3}{r} - m^2\,\phi_3=0 \ .
\ee
Thus, localized massive configurations with exponentially decaying tail may exist if the effective mass squared is positive
\be
m_{\rm eff}^2 = m^2 - g^2 V^2 > 0 \ .
\label{bound}
\ee
In the critical case $g\,V=m$ the asymptotic expansion of the charged fields $\phi_1 \pm \phi_2$ possesses a dipole term as the leading contribution, similar to the neutral mode $\phi_3$ in the massless limit, $m=0$.
The bound \re{bound} yields two different limiting cases, the electrostatic limit $g\ll V$ and the magnetic limit $g\gg V$ \cite{Livramento:2023keg}.

Unlike the previous study of $U(1)$ gauged Skyrmions in Minkowski spacetime \cite{Livramento:2023keg}, hereafter we make use of
the time dependent gauge setting $A_0(\infty)=0$.
This allows us to directly compare the properties of gauged gravitating Skyrmions and charged boson stars
\cite{Jetzer:1989us,Jetzer:1992tog,Jetzer:1991jr,Pugliese:2013gsa,Kleihaus:2009kr,Kumar:2014kna,Kunz:2021mbm}.
\\

\subsection{Axially symmetric Ansatz and boundary conditions}

Both stationary isospinning and $U(1)$ gauged configurations possess axial symmetry.
Such configurations can be parameterized by three real functions $\psi^a$ \cite{Battye:2005nx,Ioannidou:2006nn,Herdeiro:2018daq}
\begin{eqnarray}
&&\phi_1 = \psi^1(r,\theta) \cos (n \varphi-\omega t) \,, \quad \phi_2 = \psi^1(r,\theta) \sin (n \varphi-\omega t) \,,
\quad \phi_3=\psi^2(r,\theta)\,, \nonumber \\
&& \quad \phi_0 = \psi^3 (r,\theta) \ ,
\label{phi}
\end{eqnarray}
where the integer $n$ specifies the degree of the map, $B=n$, and $\omega$ can be considered as the angular frequency of the charged scalar fields.
As discussed above, the gauge freedom can be exploited to eliminate the time dependence of the charged scalar fields.
Thus in the stationary gauge and for solutions of topological degree one the Ansatz reduces to
\be
\phi_1 = \psi^1(r,\theta) \cos \varphi\,, \quad \phi_2 = \psi^1(r,\theta) \sin \varphi\,,
\quad \phi_3=\psi^2(r,\theta)\,, \quad \phi_0 = \psi^3 (r,\theta) \ .
\ee

The gauge field is parameterized by the electric and magnetic potentials $A_0$ and $A_\varphi$, respectively,
\begin{eqnarray}
\label{A}
A \equiv A_\mu dx^\mu=A_0(r,\theta) dt + A_\varphi (r,\theta) d\varphi \ .
\end{eqnarray}
Notably gauged Skyrmions are not spherically symmetric even in the sector of topological degree one \cite{Piette:1997ny,Livramento:2023keg}.
Instead the soliton is deformed by the toroidal magnetic flux in the equatorial plane.

The metric can be written in isotropic coordinates in the form
\be
ds^2 = -F_0dt^2 + F_1(dr^2 + r^2 d\theta^2)+ F_2 r^2\sin^2 \theta \left(d\varphi - \frac{W}{r} dt \right)^2 \ ,
\label{metric}
\ee
where the four functions $F_0, F_1, F_2$ and $W$ depend on $r$ and $\theta$, only.

Substitution of the Ansatz \re{phi}, \re{A} and \re{metric} yields a set of 9 coupled elliptic partial differential equations with mixed derivatives, to be solved numerically subject to appropriate boundary conditions.
These follow from the conditions of asymptotic flatness, requirements of regularity of the fields on the symmetry axis, as well as the condition of finiteness of the metric and finiteness of the $T_t^t$ and $T^t_\phi$-components of the energy-momentum tensor \re{T}.

Explicitly, for the Skyrmion of topological degree one in the time dependent gauge, we impose at the origin
\be
\begin{split}
\psi_1\big|_{r=0}&=0,~~
\psi_2\big|_{r=0}=0,~~
\psi_3\big|_{r=0}=-1,~~
A_\varphi\big|_{r=0}=0,~~
\partial_r A_0 \big|_{r=0}=0,\\
\partial_r F_0 \big|_{r=0}&=0,~~
\partial_r F_1 \big|_{r=0}=0,~~
\partial_r F_2 \big|_{r=0}=0,~~
W \big|_{r=0}=0\, ,
\end{split}
\ee
while the boundary conditions at spatial infinity are
\be
\begin{split}
\psi_1\big|_{r=\infty}&=0,~~
\psi_2\big|_{r=\infty}=0,~~
\psi_3\big|_{r=\infty}=1,~~
A_0 \big|_{r=\infty}= 0,~~
A_\varphi\big|_{r=\infty}=0,~~\\
F_0 \big|_{r=\infty}&=1,~~
F_1 \big|_{r=\infty}=1,~~
F_2 \big|_{r=\infty}=1,~~
W \big|_{r=\infty}=0\,.
\end{split}
\label{BC_infty}
\ee
The condition $\partial_r A_0(0)=0$ ensures  the electric field to be absent at the center of configuration.

Finally, to ensure the condition of regularity on the symmetry axis we impose the boundary conditions
\be
\begin{split}
\psi_1\big|_{\theta=0,\pi}&=0,~~
\partial_\theta\psi_2\big|_{\theta=0,\pi}=0,~~
\partial_\theta\psi_3\big|_{\theta=0,\pi}=1,~~
\partial_\theta A_0 \big|_{\theta=0,\pi}=0,~~
A_\varphi\big|_{\theta=0,\pi}=0,~~\\
\partial_\theta F_0 \big|_{\theta=0,\pi}&=0,~~
\partial_\theta F_1 \big|_{\theta=0,\pi}=0,~~
\partial_\theta F_2 \big|_{\theta=0,\pi}=0,~~
\partial_\theta W \big|_{\theta=0,\pi}=0\,.
\end{split}
\ee
In addition, requiring the absence of a conical singularity on the symmetry axis demands that the deficit angle should vanish, i.e., $\delta= 2\pi(1-\lim_{\theta\to 0}({F_2}/{F_1})=0)$.
Hence any physically consistent solution should satisfy the constraint ${F_1}_{\theta=0}={F_2}_{\theta=0}$.
In our numerical scheme we explicitly impose this condition on the symmetry axis.

We have solved the boundary value problem subject to the boundary conditions above with a sixth-order finite difference scheme, where the system of equations is discretized on a grid with a typical
size of $229\times 79$ points.
The corresponding system of nonlinear algebraic equations has been solved using the Newton-Raphson scheme.
Calculations have been performed with the packages FIDISOL/CADSOL \cite{Schoenauer:1989,Schoenauer:1989b}, with typical errors of order of $10^{-4}$.

\subsection{Truncation to $O(3)$ Faddeev-Skyrme model}

Non-topological soliton solutions are found for a consistent truncation of the Skyrme model to the $O(3)$ Faddeev-Skyrme model \cite{Ioannidou:2006nn,Perapechka:2017bsb,Herdeiro:2018daq} with (cf \re{phi})
\be
\phi_1= \psi_1(r,\theta) \cos (n \varphi - \omega t)\,, \quad \phi_2 = \psi_1(r,\theta) \sin (n \varphi - \omega t)\,,
\quad \phi_3=0\,, \quad \phi_0 = \psi_3 (r,\theta)
\label{phi-sigma}
\ee
and the sigma-model constraint $\psi_1^2+\psi_3^2=1$.
Hence, it is convenient to parameterize the fields as $\psi_1(r,\theta)= \sin(h),~~\psi_3(r,\theta)= \cos(h)$, where $h(r,\theta)$ is the profile function.
Solutions for $n=1$ obey the same set of boundary conditions as imposed for the Skyrmion of topological degree one above \cite{Ioannidou:2006nn}.

However, for $n=0$ the solutions possess spherical symmetry.
In this case we parametrize the metric in Schwarzschild-like coordinates, for convenience,
\be
ds^2 = -\sigma^2(r) N(r) dt^2 + \frac{1}{N(r)}dr^2 + r^2 d\Omega^2 \ .
\label{metric-Schwarz}
\ee
In terms of this parametrization, the Skyrme-Maxwell Lagrangian \re{Lag0b} in the time dependent gauge becomes \footnote{Coordinate transformations from the axially symmetric isotropic metric \re{metric} to the Schwarzschild-like line element \re{metric-Schwarz} are discussed, for example, in \cite{Hartmann:2001ic}.}
\be
{\cal L}_{m}=\frac{\left(A_0^\prime\right)^2}{\sigma^2}+\frac{(\omega - gA_0)^2 \sin^2h}{\sigma^2 N} - N ~{h^\prime}^2 + \frac{(\omega - gA_0)^2 \sin^2h ~{h^\prime}^2}{\sigma^2}-m^2 (1-\cos h) \ ,
\ee
and the reduced curvature scalar is
\be
R=-2\sigma (N-1 +r N^\prime) \ ,
\ee
where a prime denotes the radial derivative.
Thus, the frequency appears in the field equations only in the combination $\omega_g=\omega - g A_0$.
The effective mass squared is now given by
\be
m_{\rm eff}^2=m^2 - (\omega - g A_0)^2
\ \ \ \text{with} \ \ \
\lim\limits_{r \rightarrow \infty}{m_{\rm eff}^2} = m^2 - \omega^2 \ ,
\ee
and localized solutions exist only if $m^2 - \omega^2 > 0$.
Note that, in the time dependent gauge, this
bound exactly matches the analogous upper mass threshold for asymptotically flat boson stars and Q-balls \cite{Volkov:2002aj,Kleihaus:2005me,Collodel:2019ohy}.

For the spherically symmetric solutions we employ the following set of the boundary conditions
\be
\begin{split}
\partial_r\sigma\big|_{r=0}&=0,~~
N\big|_{r=0}=1,~~
\partial_r h\big|_{r=0}=0 , \\
\sigma \big|_{r=\infty}&=1,~~
N \big|_{r=\infty}=1,~~
h \big|_{r=\infty}=0 .
\end{split}
\ee

\subsection{Physical properties}

The ADM mass $M$ of the solutions can be read off from the asymptotic subleading behavior of the metric function $g_{00}$,
\be
g_{00}(r)= -1 + \frac{\alpha^2 M}{\pi r}+\dots \ .
\ee
Similarly, the electric charge of the gauged Skyrmions can be computed from the far field expansion of the electric potential,
\be
A_0(r)= \frac{Q}{r} + O\left(\frac{1}{r^2}\right) \ .
\ee
The angular momentum $J$ can be read off from the asymptotic behavior of the metric function $g_{0\varphi}$,
\be
g_{0\varphi}=\frac{\alpha^2 J}{\pi r}\sin^2\theta+ O\left(\frac{1}{r^2}\right) \ .
\label{J}
\ee
The angular momentum $J$ and the electric charge $Q$ are proportional, $J \propto n Q$ \cite{Radu:2008pp,Herdeiro:2019mbz,Livramento:2023keg}. Both quantities can also be computed as the integrals of the corresponding components of the total stress-energy tensor $T^{\mu\nu}$, eq.~\re{T}.

Finally, the magnetic dipole moment $\mu_{m}$ can be computed from the far field expansion of the magnetic potential,
\be
A_\varphi = \frac{\mu_m}{r}\sin^2 \theta+O\left(\frac{1}{r^2}\right)
 \ .
\ee
\\

\section{Gravitating gauged Skyrmions}

The usual flat space $B=1$ Skyrmion solution is recovered in the flat space limit $\alpha=0$, when the electromagnetic field is decoupled $g=0$.
From this Skyrmion a branch of gauged flat space Skyrmions arises, when the gauge coupling is increased from zero.
Since the field equations impose the condition $\omega = |gV| \le m$, in order to obtain localized Skyrmion solutions, the maximal value of the coupling constant $g$ is restricted by this condition \cite{Livramento:2023keg}.
In our numerical simulations below we set $m=1$ and fix $g=0.1$. This choice matches the related recent study of gauged boson stars in the two-component scalar model \cite{Kunz:2021mbm}.

Self-gravitating generalizations of the gauged Skyrmions are found by increasing gradually the effective gravitational coupling constant $\alpha$ for a given value of $g$.
In the time dependent gauge the angular frequency $\omega$ is an input parameter.
Setting $\omega=0$ corresponds to the solutions with zero electric charge.
First, we observe that for all values of the parameters, a branch of gravitating electrically charged solutions emerges from the corresponding flat space configuration, as the effective gravitational coupling $\alpha$ increases from zero.
All these solutions are coupled to a local magnetic flux, which generates a non-vanishing magnetic moment.
Therefore the electromagnetic interaction breaks spherical symmetry of the $B=1$ Skyrmion.

The dependence of the ungauged Skyrmions on gravity has been studied before \cite{Glendenning:1988qy,Bizon:1992gb,Heusler:1991xx,Heusler:1993ci}.For any allowed value of the effective gravitational coupling $\alpha^2=\frac{1}{2}\pi G f^2_\pi$ there are two branches of solutions which merge at a maximal value of $\alpha$, $\alpha_{cr}$.
The branch of solutions lower in energy is linked to the flat space Skyrmions in the limit $\alpha\to 0$, i.e., when $G \to 0$.
At $\alpha_{cr}$ it bifurcates with the second branch of solutions, that is higher in energy.
This second branch extends backward towards a limiting strongly gravitating solution.
Here the limit $\alpha\to 0$ is approached as $f_\pi\to 0$, and corresponds to the absence of the quadratic term in the Skyrme Lagrangian.
As shown in \cite{Bizon:1992gb}, in this limit the configuration  approaches the lowest mass spherically symmetric Bartnik-McKinnon solution \cite{Bartnik:1988am}.

The contribution of the electromagnetic energy slightly modifies this pattern.
The spike of the uncharged case transforms into a loop (see e.g., \cite{Kleihaus:2007vf}) and the value of $\alpha_{cr}$ increases as $\omega$ grows.
This is illustrated in Fig.~\ref{fig1}.
The breaking of spherical symmetry of the configurations on the fundamental branch is maximal as $\omega$ approaches the mass threshold and the gravitational coupling $\alpha$
remains relatively small.

\begin{figure}[t!]
\begin{center}

\includegraphics[height=.33\textheight,  angle =-90]{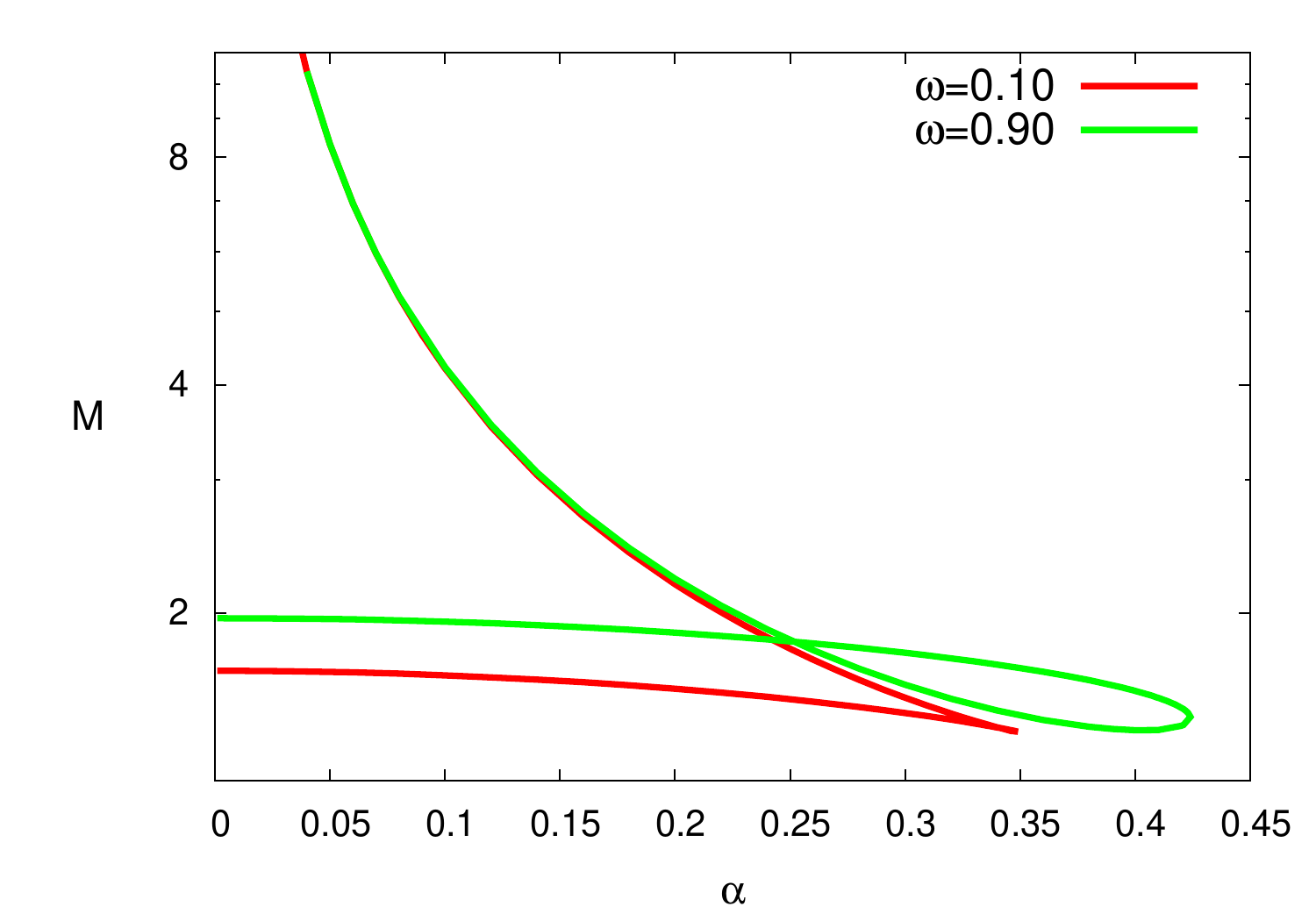}
\includegraphics[height=.33\textheight,  angle =-90]{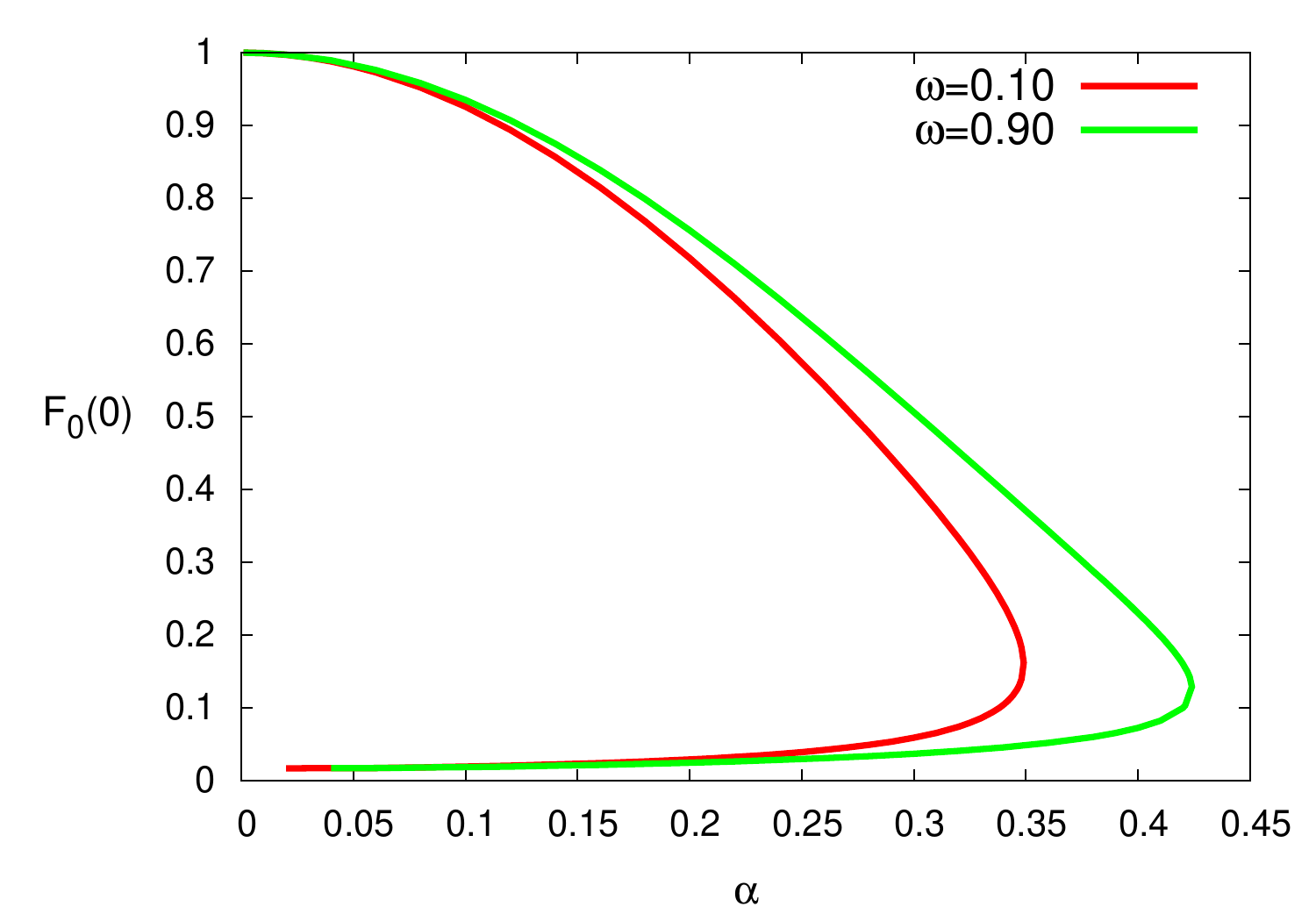}
\includegraphics[height=.33\textheight,  angle =-90]{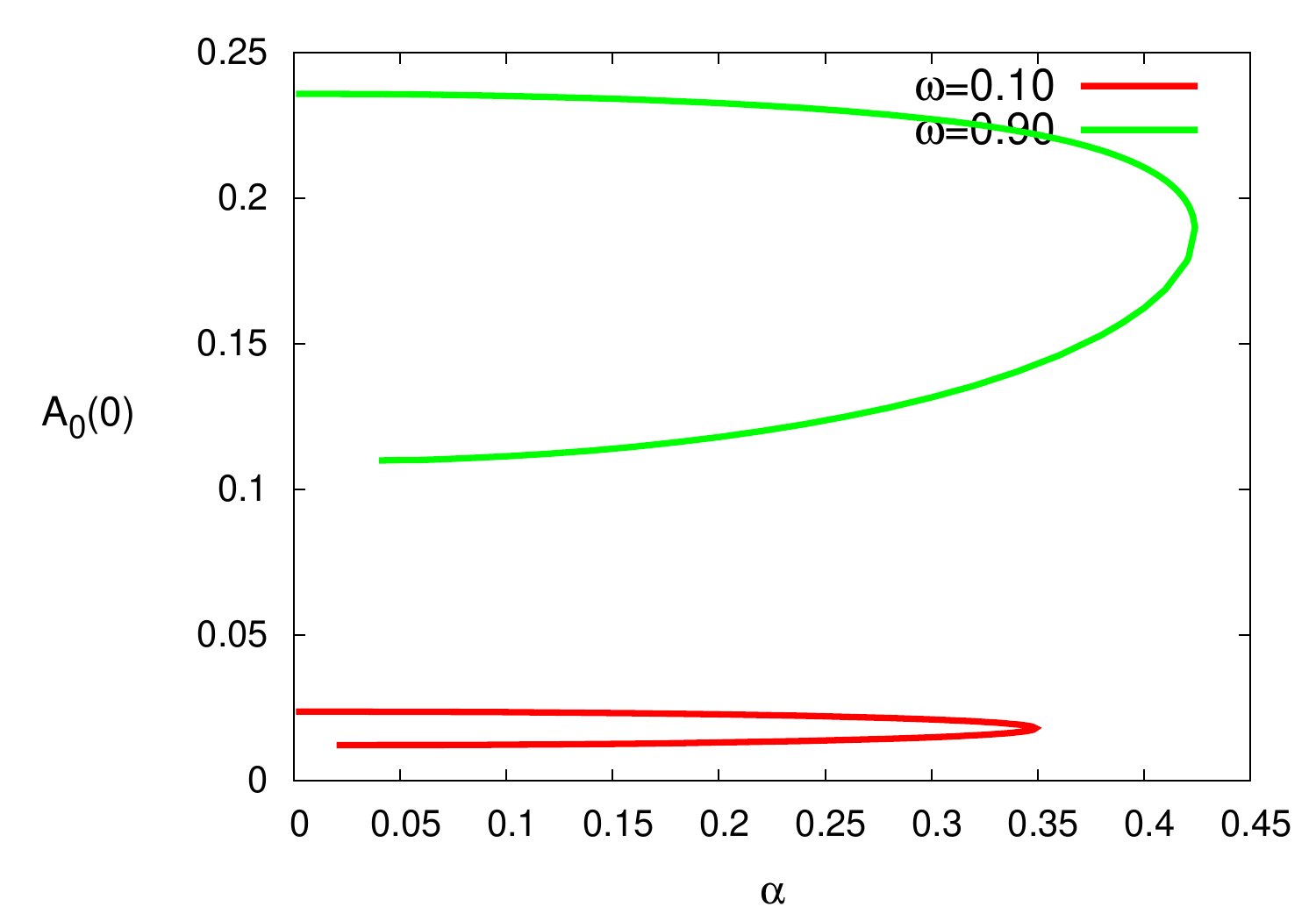}
\includegraphics[height=.33\textheight,  angle =-90]{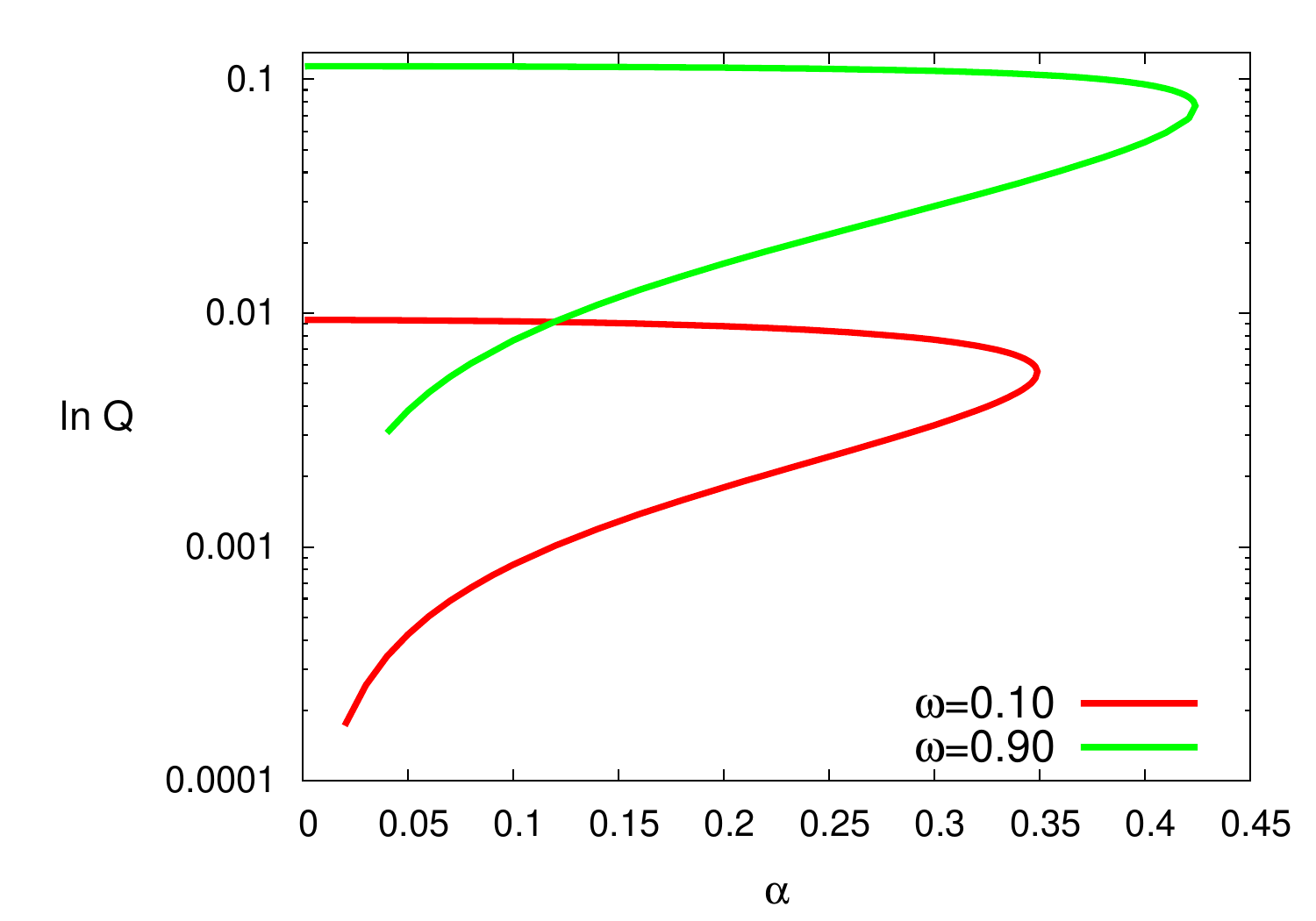}
\includegraphics[height=.33\textheight,  angle =-90]{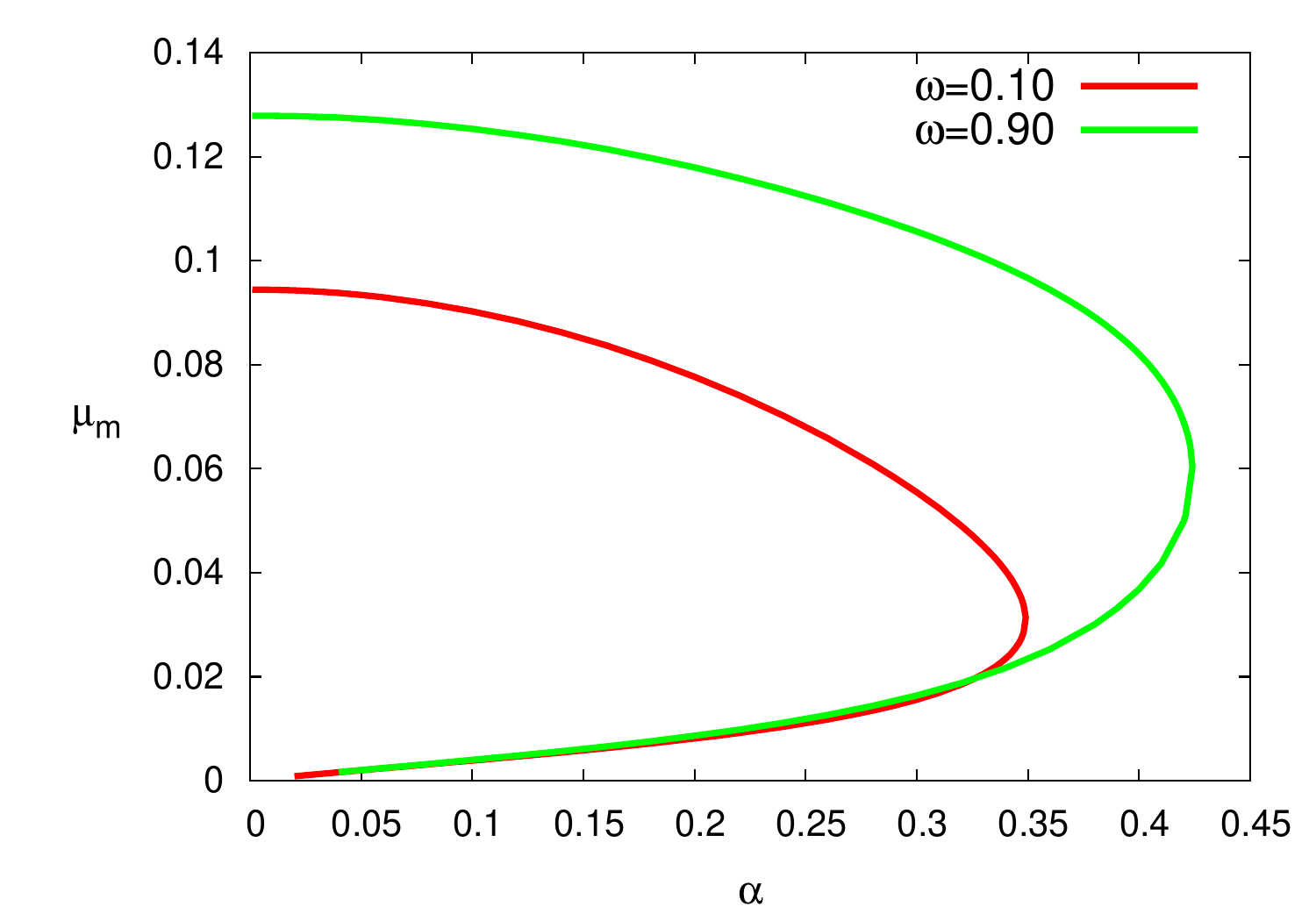}

\end{center}
\caption{\small Gravitating gauged Skyrmions: The mass $M$ (upper left plot), the value of the metric function $F_0$ at the origin (upper right plot), the value of the gauge potential $A_0$ at the origin (middle left plot), the charge $Q$ (middle right plot), and the magnetic moment $\mu_m$ of the configurations (lower plot) are displayed as functions of the effective gravitational coupling $\alpha$ for $g=0.1,\, m=1$ and $\omega=0.1, \, 0.9$.}
    \lbfig{fig1}
\end{figure}

However, for the configurations on the second branch the effect of the electromagnetic interaction becomes less and less
important as $\alpha$ is decreasing.
One can understand the reason of this when introducing the rescaled radial coordinate $\tilde r= r/\alpha$ \cite{Bizon:1992gb}.
Then the covariant derivative of the Skyrme field rescales as
$\widetilde D_\mu \phi^a = \alpha D_\mu \phi^a$ with $\tilde A_\mu=\alpha A_\mu$.
Evidently, the contribution of the Maxwell field in the limiting $\alpha\to 0$ configuration is vanishing.

\begin{figure}[t!]
\begin{center}

\includegraphics[height=.33\textheight,  angle =-90]{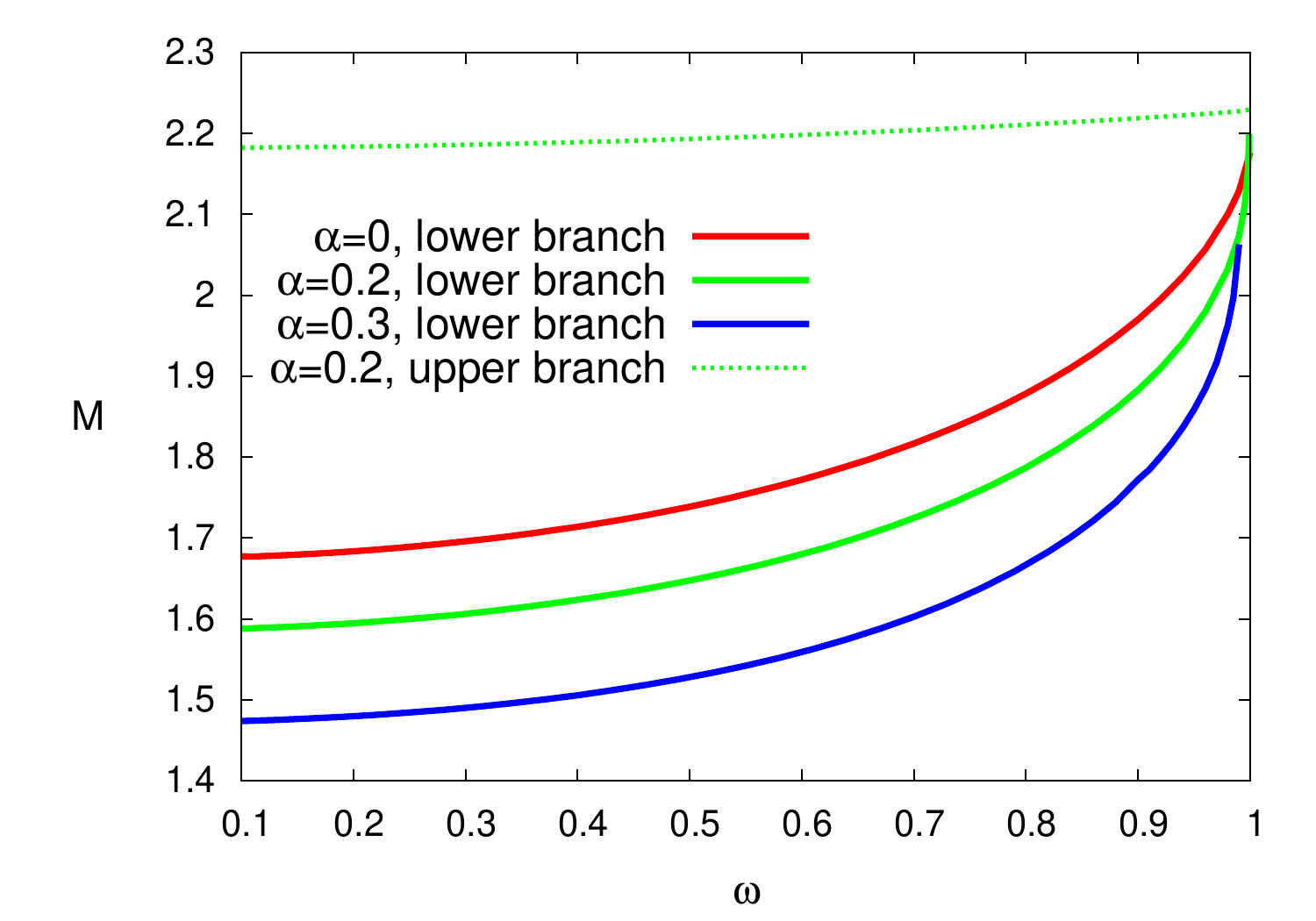}
\includegraphics[height=.33\textheight,  angle =-90]{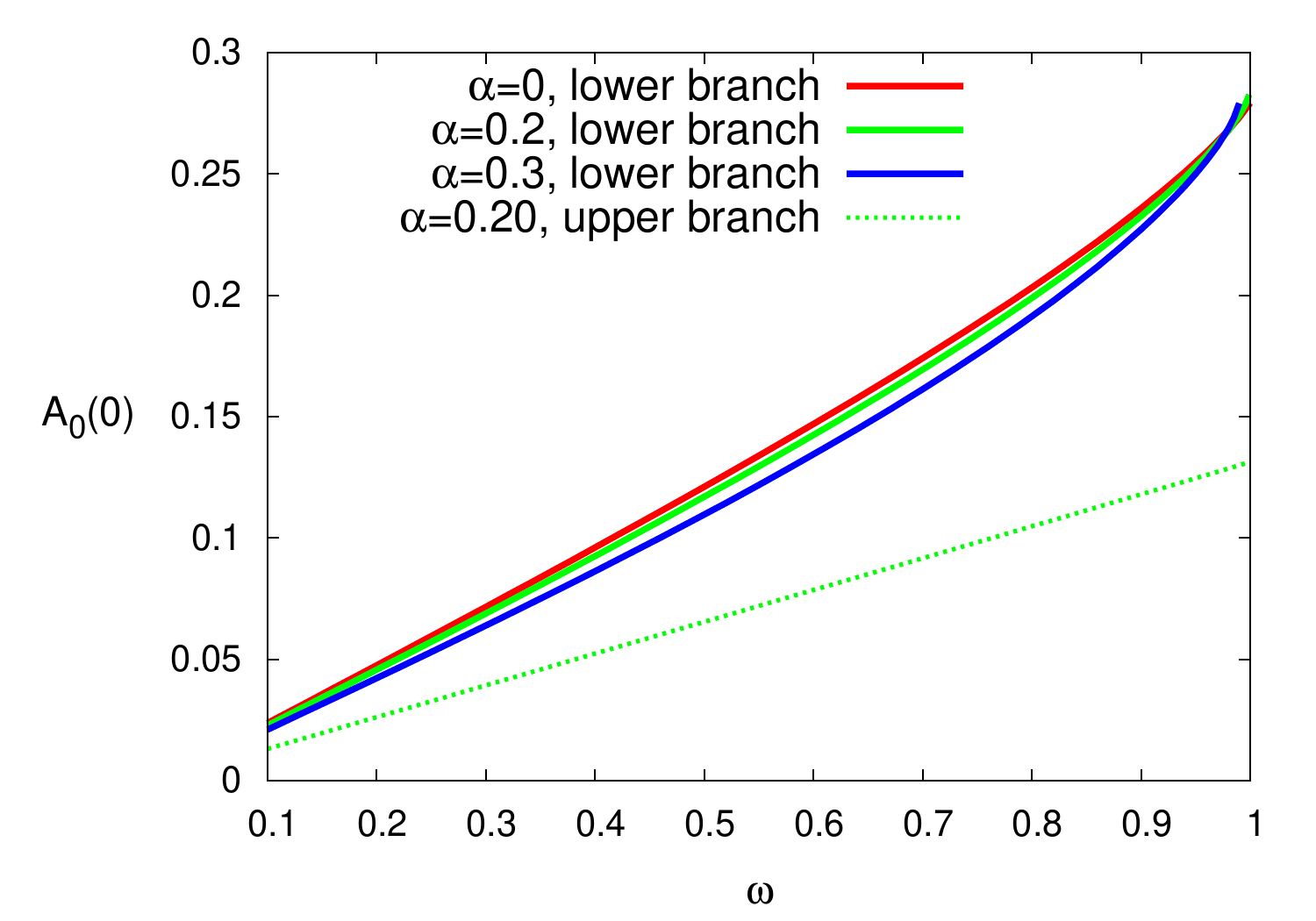}
\includegraphics[height=.33\textheight,  angle =-90]{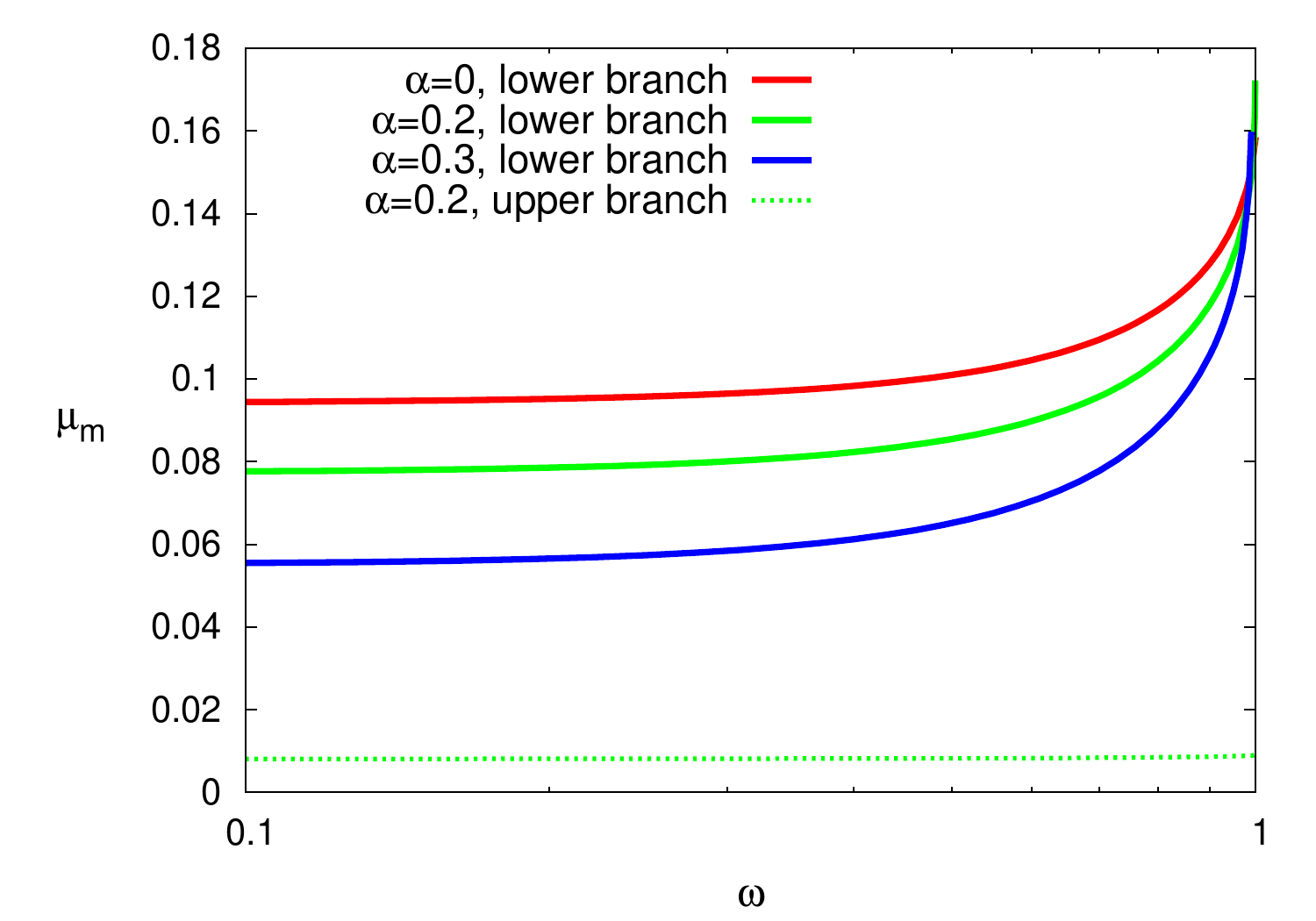}
\includegraphics[height=.33\textheight,  angle =-90]{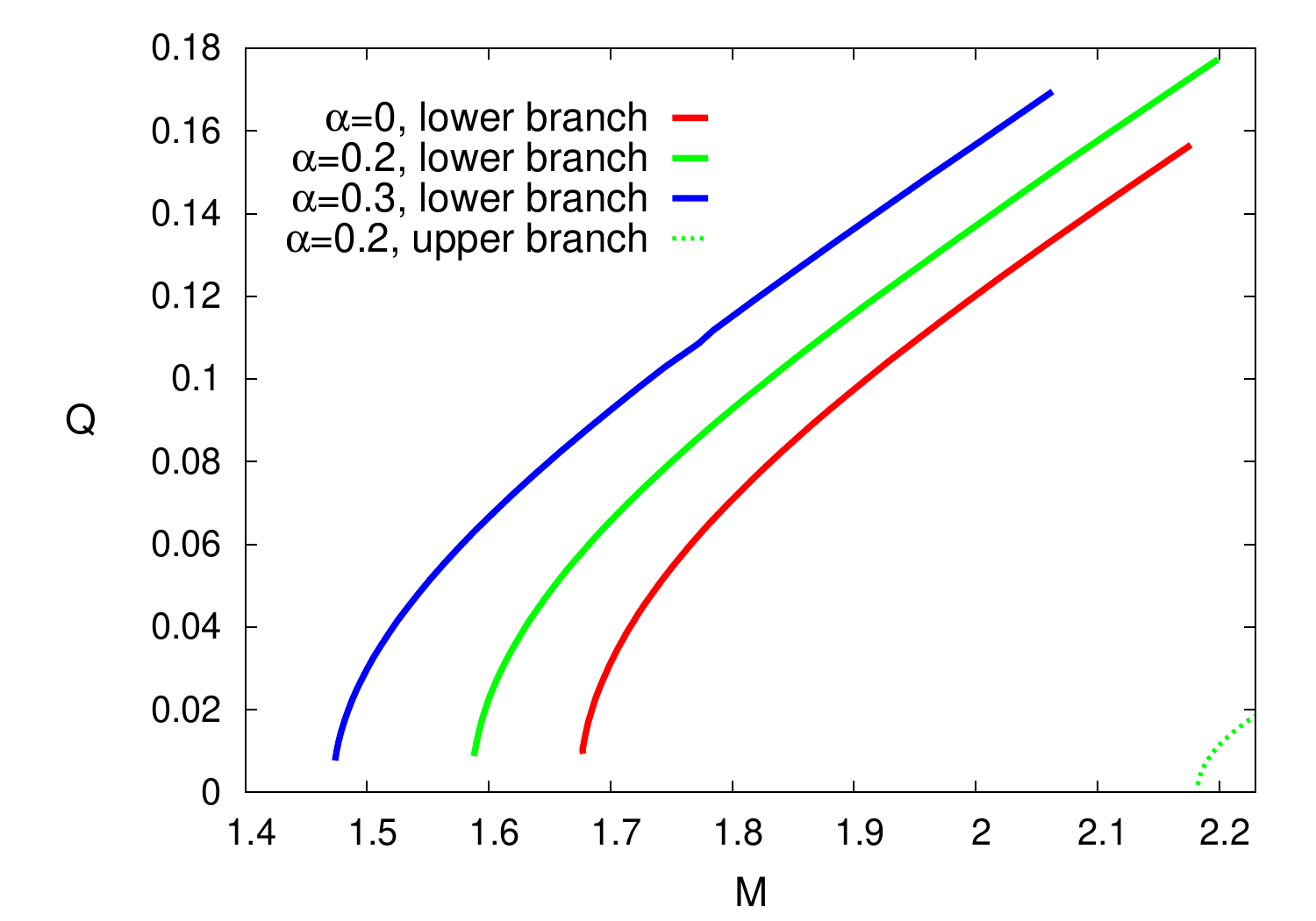}

\end{center}
\caption{\small Gravitating gauged Skyrmions: The mass $M$ (upper left plot), the value of the gauge potential $A_0$ at the origin (upper right plot), and the magnetic moment $\mu_m$ (bottom left plot) of the configurations are plotted as functions of the frequency $\omega$ for $g=0.1,\, m=1$ and several values of the gravitational coupling $\alpha$.
The ADM mass $M$ of these solutions vs the charge $Q$ is displayed in the bottom right plot.}
    \lbfig{fig2}
\end{figure}

Considering the dependence of the solutions on the frequency $\omega$ we observe monotonic increase of the ADM mass and the charge with increasing $\omega$ and a corresponding gain of electromagnetic energy for all regular self-gravitating Skyrmions, as seen in Fig.~\ref{fig2}.
\\


\section{Pion stars}

\subsection{Uncharged pion clouds ($g=0$)}

Pion clouds are obtained in the truncated Skyrme model, where $\phi_3=0$ (see eq.~\re{phi-sigma}).
Apart from the sigma-model constraint these configurations are akin to Q-balls.
They exist for some limited range of values of parameters of the model \cite{Ioannidou:2006nn,Perapechka:2017bsb,Herdeiro:2018daq}.
It has been shown that such axially-symmetric spinning configurations dubbed as \textit{pion clouds}, exist for $n=1$.

Surprisingly, for $n=0$ also spherically symmetric clouds appear to exist for sufficiently large values of the frequency $\omega$ due to the force balance between the gravitational and scalar interactions, both for electrically charged configurations and in the absence of electrostatic repulsion.
In the Einstein-Skyrme model we identify these solutions with  \emph{pion stars} \cite{Brandt:2018bwq}.

The appearance of pion stars (or pion clouds) can be explained by a certain balance condition that involves the repulsive force from the kinetic term that is quadratic in derivatives in \re{Lag}, and the attractive interaction from the quartic Skyrme term in the curved spacetime.
Indeed, the effective gravitational coupling in the Einstein-Skyrme model $\alpha^2=\frac{1}{2}\pi G f_\pi^2$ depends both on the Newton constant $G$ and the pion decay constant $f_\pi$.
However, unlike the topologically non-trivial Skyrmion solutions considered above, there is no branch of pion star solutions linked to the flat space limit.
On the single branch in the $(\alpha, M)$ diagram both the mass and the charge of the solutions increase with decreasing $\alpha$.
This pattern is similar to the case of the evolution of the self-gravitating Skyrmions on the second, higher energy branch, as illustrated in Fig.~\ref{fig3}.

In the absence of the electromagnetic interaction ($g=0$) the branch of pion star solutions extends all the way down to the limiting solution at $\alpha\to 0$ \cite{Ioannidou:2006nn}.
In the limiting case $\alpha \to \infty$, the quartic Skyrme term in the matter field Lagrangian \re{Lag0b} becomes negligible and the system is effectively reduced to the usual Einstein-Klein-Gordon model supporting boson star solutions.
In the opposite limit $\alpha\to 0$ the Skyrme term dominates and the resulting solution rapidly extends over the whole space \cite{Ioannidou:2006nn}.

\begin{figure}[t!]
\begin{center}

\includegraphics[height=.33\textheight,  angle =-90]{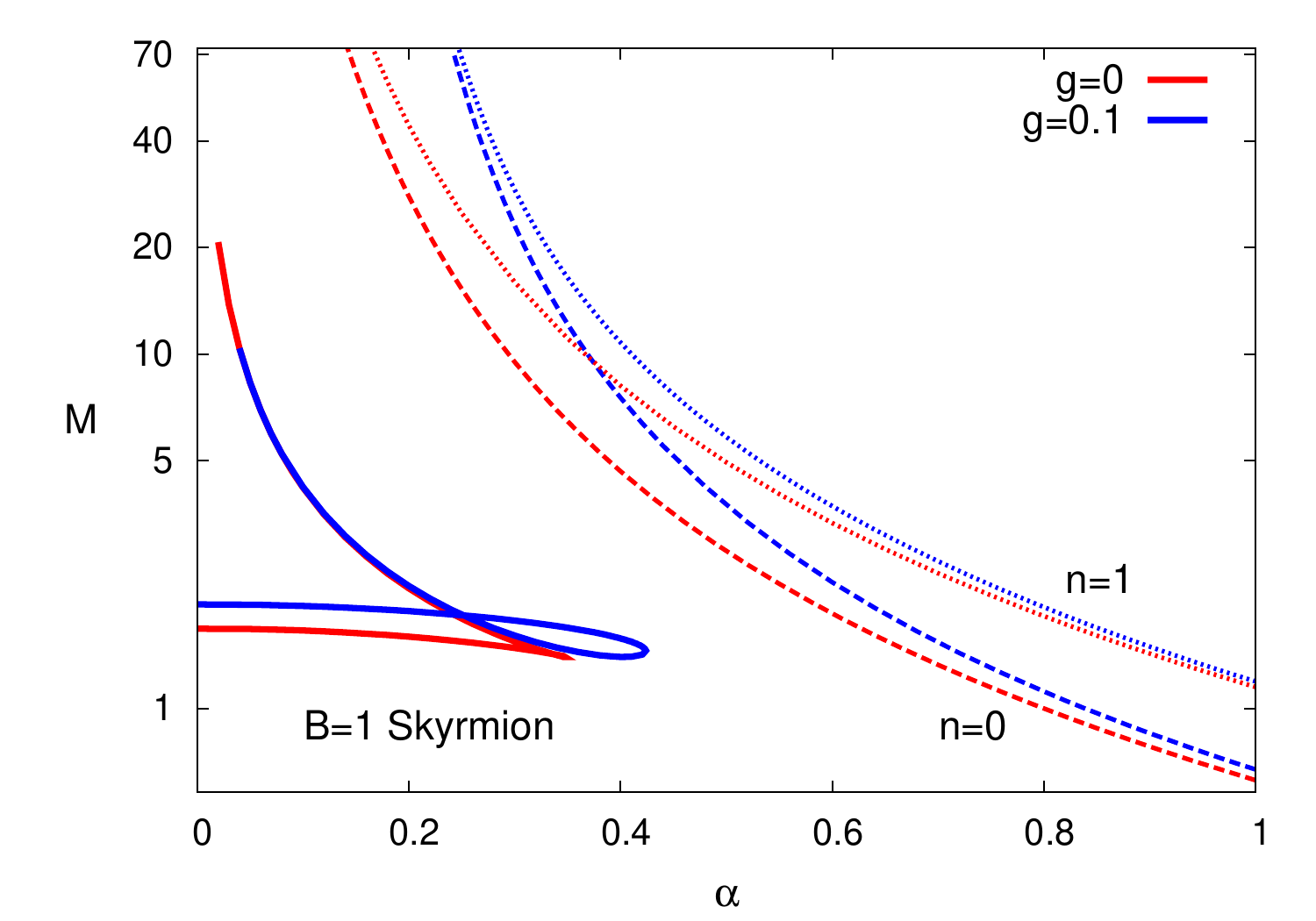}
\includegraphics[height=.33\textheight,  angle =-90]{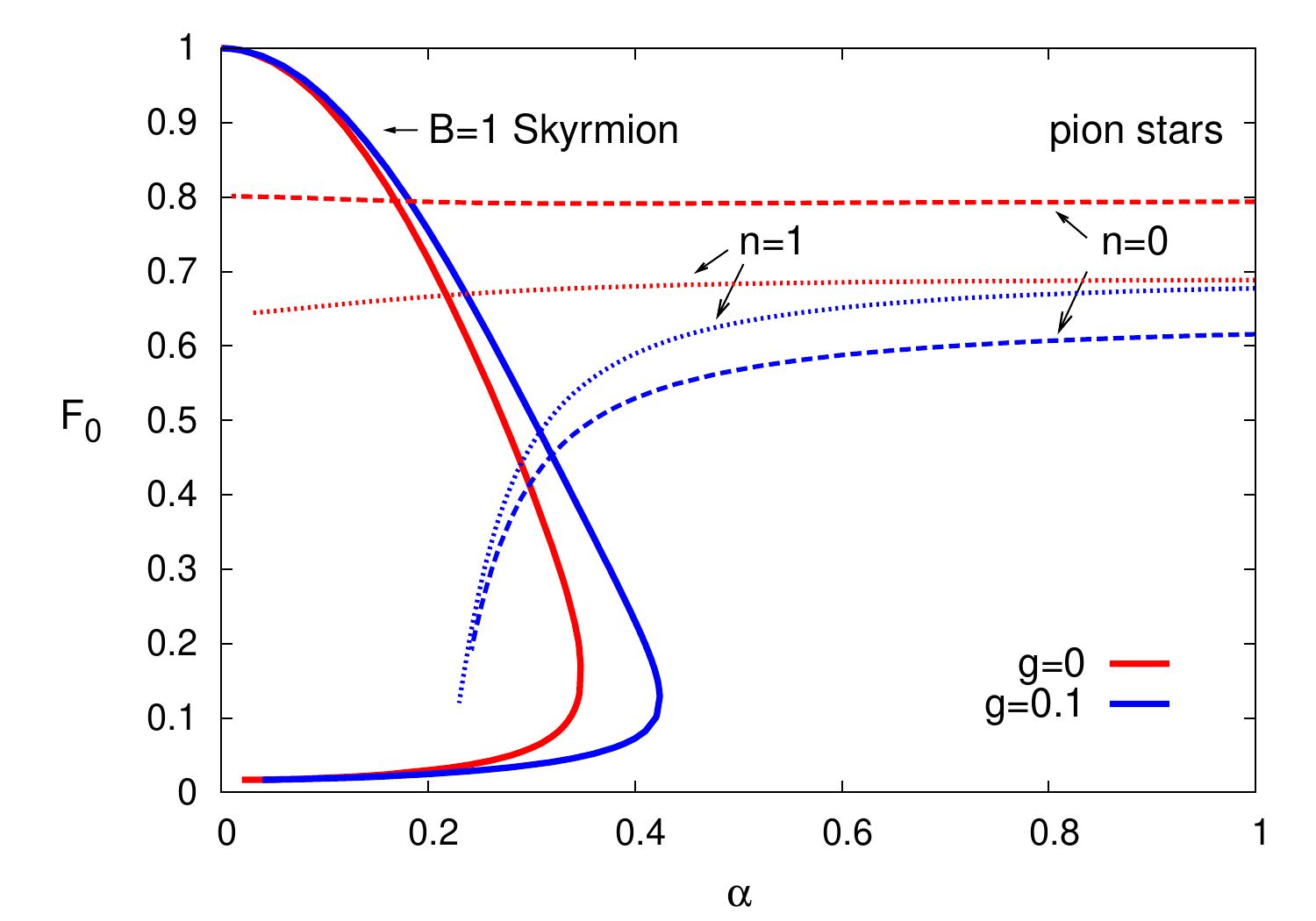}
\end{center}
\caption{\small Self-gravitating pion stars ($n=0$) compared to Skyrmions: The mass $M$ (left plot), and the value of the metric function $F_0$ at the origin (right plot) are displayed as functions of the effective gravitational coupling $\alpha$ for $g=0$ (ungauged limit) and $g=0.1,\, m=1$ and $\omega=0.9$.}
    \lbfig{fig3}
\end{figure}
\begin{figure}[tbh]
\begin{center}
\includegraphics[height=.33\textheight,  angle =-90]{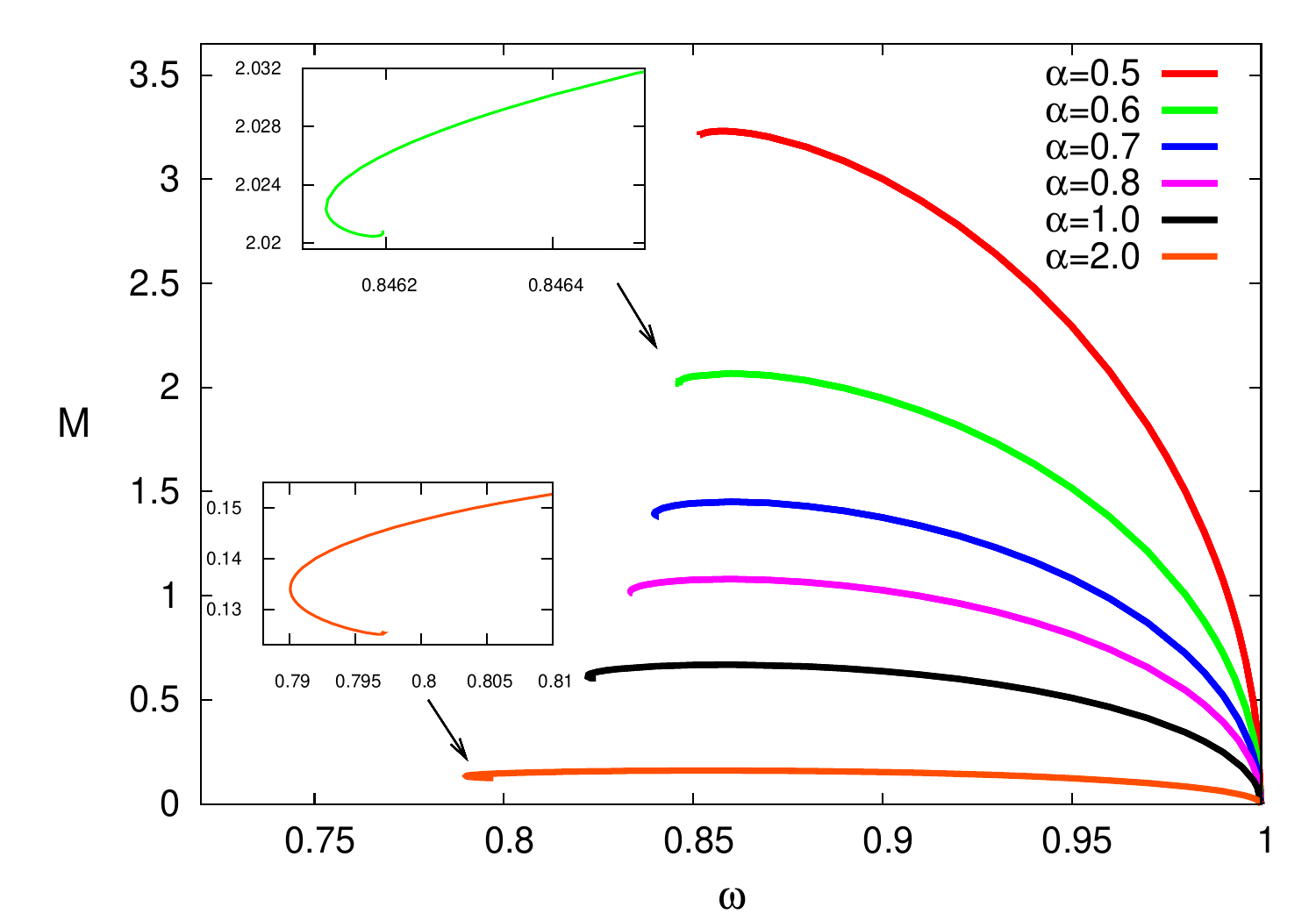}
\includegraphics[height=.33\textheight,  angle =-90]{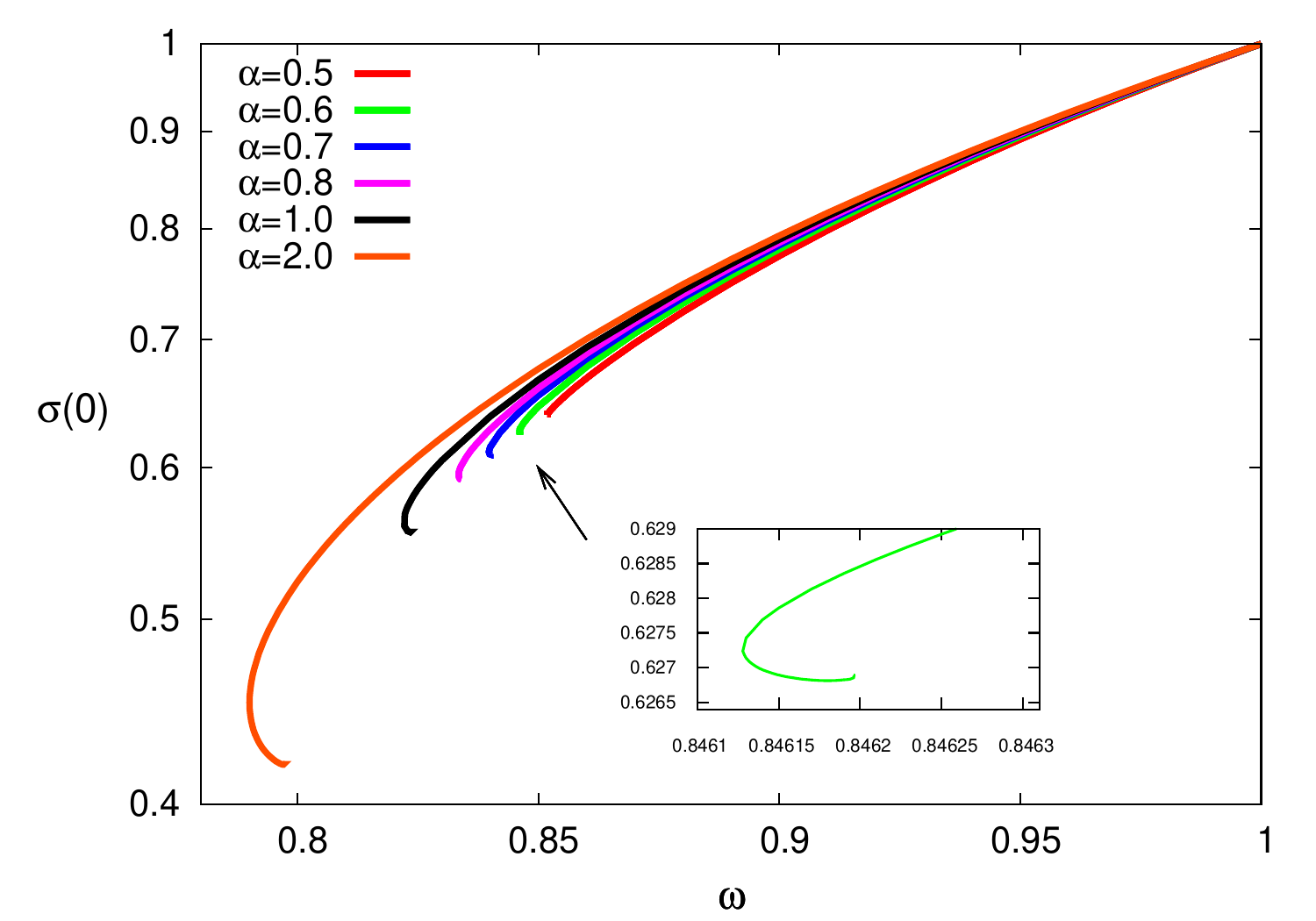}
\includegraphics[height=.33\textheight,  angle =-90]{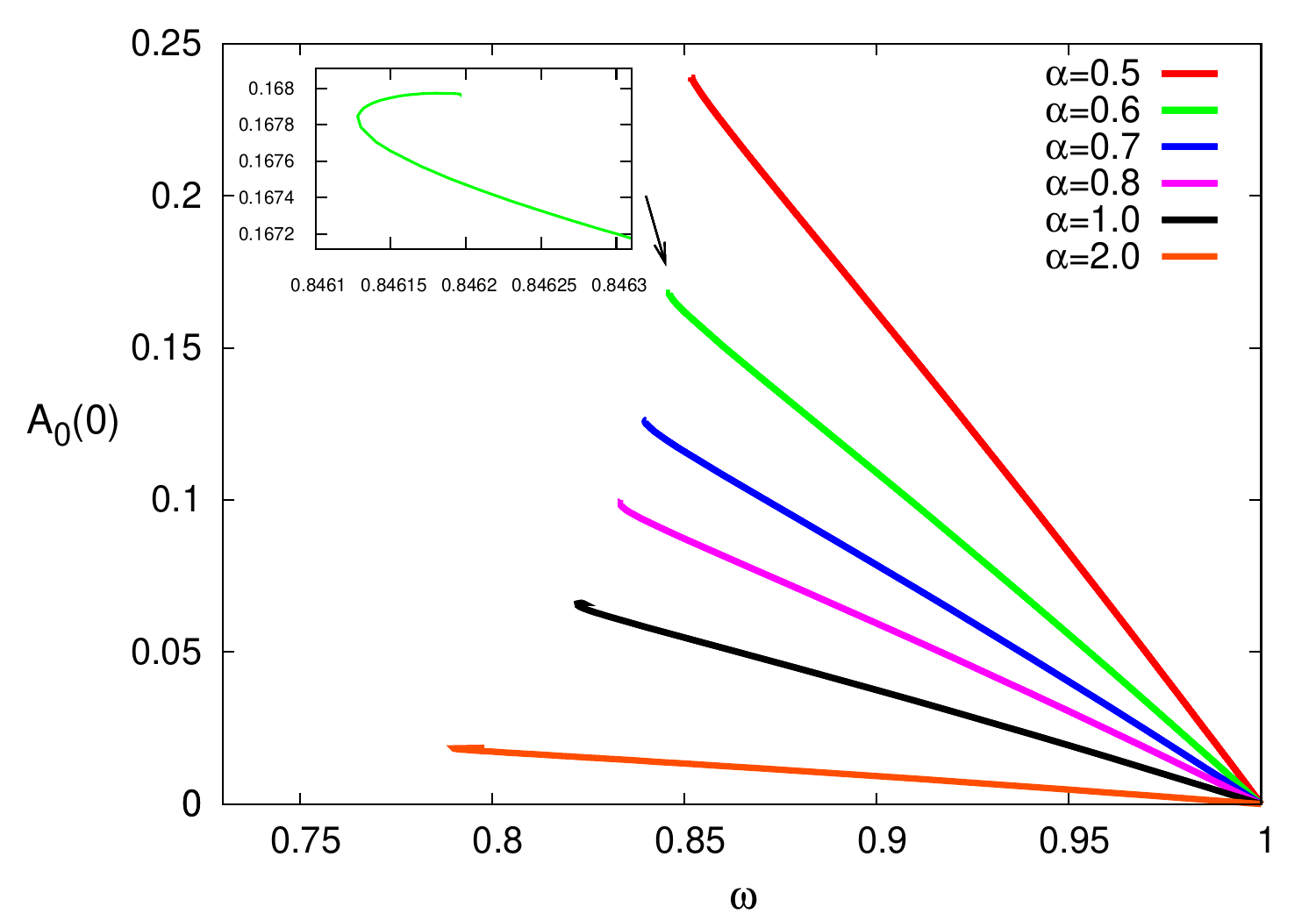}
\includegraphics[height=.33\textheight,  angle =-90]{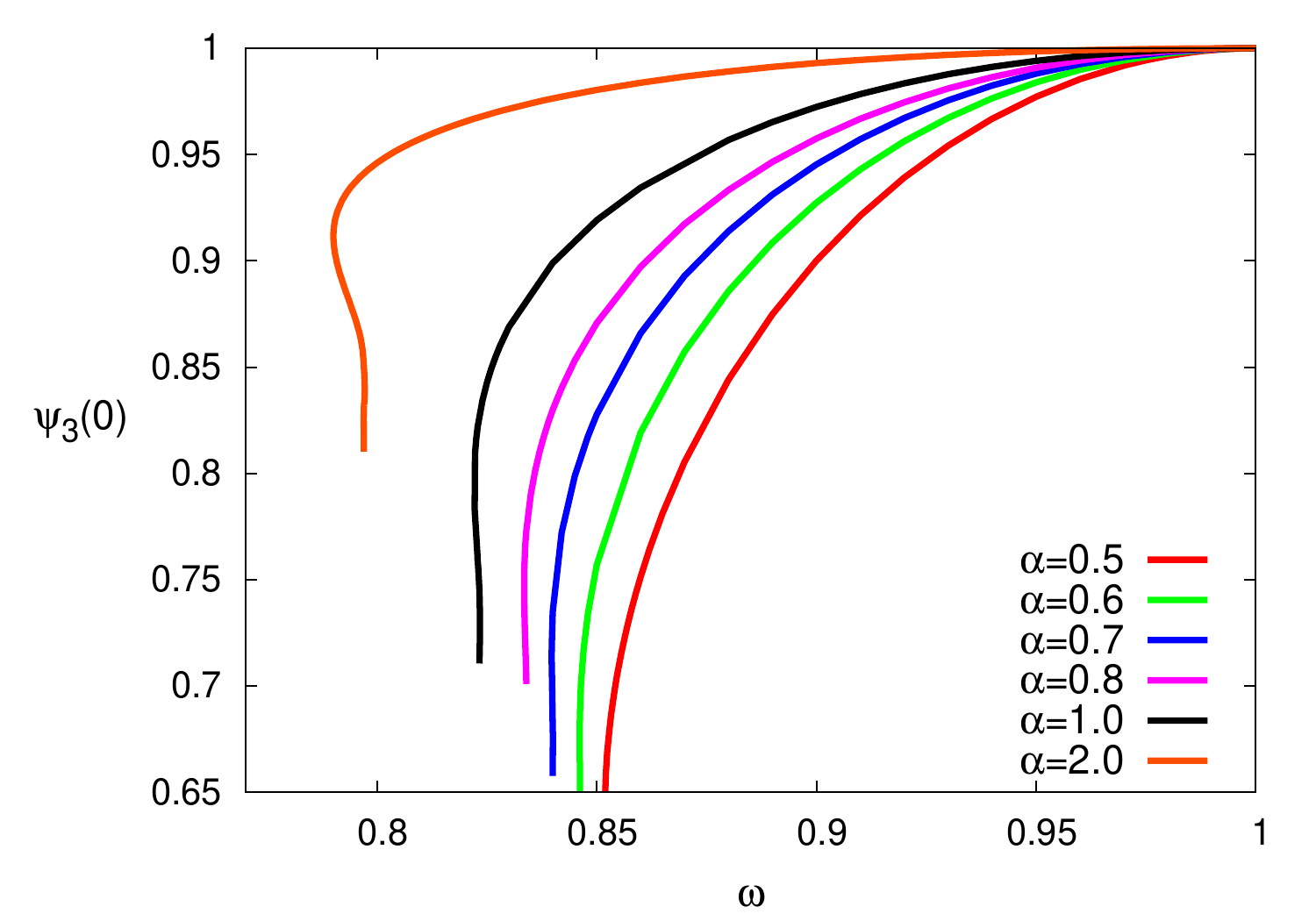}
\end{center}
\caption{\small Pion stars ($n=0$): The mass $M$ (upper left plot),
the value of the metric function $\sigma$ at the origin (upper right plot), the value of the gauge potential $A_0$ at the origin (bottom left plot), and the value of the scalar field $\psi_3$ at the origin (bottom right plot) are displayed as functions of the frequency $\omega$ for $g=0.1,\, m=1$ and several values of $\alpha$.
The insets show a magnification of the transition to the second branch.
}
    \lbfig{fig4}
\end{figure}
\begin{figure}[t!]
\begin{center}
\includegraphics[height=.40\textheight,  angle =-90]{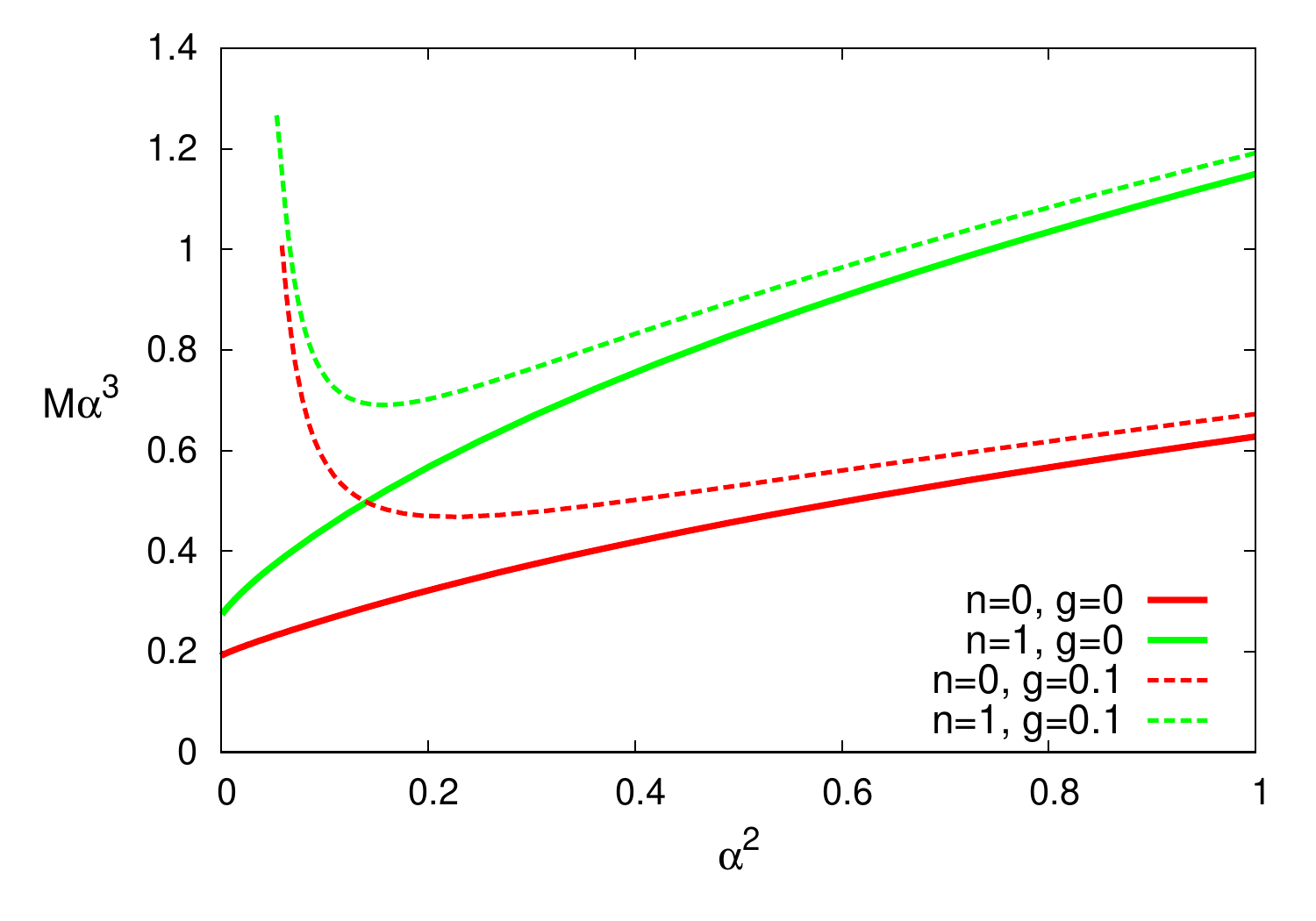}
\end{center}
\caption{\small Self-gravitating pion stars: The scaled mass $M \alpha^3$ of the $n=0,1$  solutions vs   $\alpha^2$ for $g=0$ (ungauged limit) and $g=0.1,\, m=1$ and $\omega=0.9$.}
    \lbfig{fig5}
\end{figure}

One notices that, similar to the Skyrmions, for a given $\alpha$ the solutions exist for $\omega \le m$.
When $\omega$ is decreased from the pion mass threshold, while keeping $\alpha$ and the other parameters fixed, a fundamental branch of gravitating pion star solutions emerges from the vacuum in the curved spacetime.

Considering the dependence of the pion stars on the angular frequency we observe the pattern familiar from the corresponding study of the boson stars \cite{Jetzer:1989us,Jetzer:1992tog,Jetzer:1991jr,Pugliese:2013gsa,Kleihaus:2009kr,Kumar:2014kna,Kunz:2021mbm}.
As the frequency is decreased below the mass threshold, both the ADM mass $M$ and the charge $Q$ reach monotonically a maximum value, as displayed in Fig.~\ref{fig4}.
Further decreasing of $\omega$ leads to a minimal frequency $\omega_{cr}$ below which no pion stars are found.
The lower energy branch of the pion stars bifurcates at $\omega_{cr}$ with a secondary branch of solutions, where $\omega$ is increasing, see Fig.~\ref{fig4}.
We conjecture that by analogy with the usual boson stars, both the $(\omega, M)$ and $(\omega, Q)$ curves follow a spiraling/oscillating pattern toward a singular limiting solution, with successive backbendings.

Note that the size of the spiral can be very small: a tiny variation of the angular frequency can strongly affect the fine force balance, as illustrated in Fig.~\ref{fig4}.
This observation is related to the difference between the mechanisms of the formation of a spiral or of damped oscillations, in the dynamical evolution of boson and pion stars, respectively.
In the former case, the appearance of the frequency-mass spiral is due to oscillations in the force balance between the repulsive scalar interaction and the gravitational attraction \cite{Jetzer:1991jr}.
For the pion stars, the spiraling/oscillating pattern may be related to the competing roles of the terms in Skyrme Lagrangian \re{Lag}, which are quadratic and quartic in derivatives of the scalar field, in the presence of a strong gravitational attraction.
Indeed, an increase of the effective gravitational coupling $\alpha$, that is related with a growing contribution of the repulsive force mediated by the quadratic term in \re{Lag}, leads to a decrease of $\omega_{cr}$.
At the same time also the secondary branches become more extended, as seen in Fig.~\ref{fig4}.\\

The $n=1$ axially symmetric pion stars in the ungauged limit were discussed in \cite{Ioannidou:2006nn,Perapechka:2017bsb}.
These spinning configurations possess angular momentum $J$ \re{J}, which, analogous to the usual boson stars, is proportional to the Noether charge $Q$ \cite{Radu:2008pp,Herdeiro:2019mbz}.

The mass of the $n=1$ axially symmetric pion stars is higher than the mass of the corresponding $n=0$ solutions for the same values of the parameters, see Figs.~\ref{fig5} and \ref{fig6}.
Also the minimal value of the frequency $\omega_{cr}$ is smaller than for the spherically symmetric pion stars, as shown in Fig.~\ref{fig6}.
In Fig.~\ref{fig5} we exhibit the scaled mass $M\alpha^3$ of the pion stars as a function of $\alpha^2$ at $\omega=0.9$ (cf. corresponding plot 5 in \cite{Ioannidou:2006nn}).

\subsection{Gauged pion stars}
\begin{figure}[t!]
\begin{center}
\includegraphics[height=.33\textheight,  angle =-90]{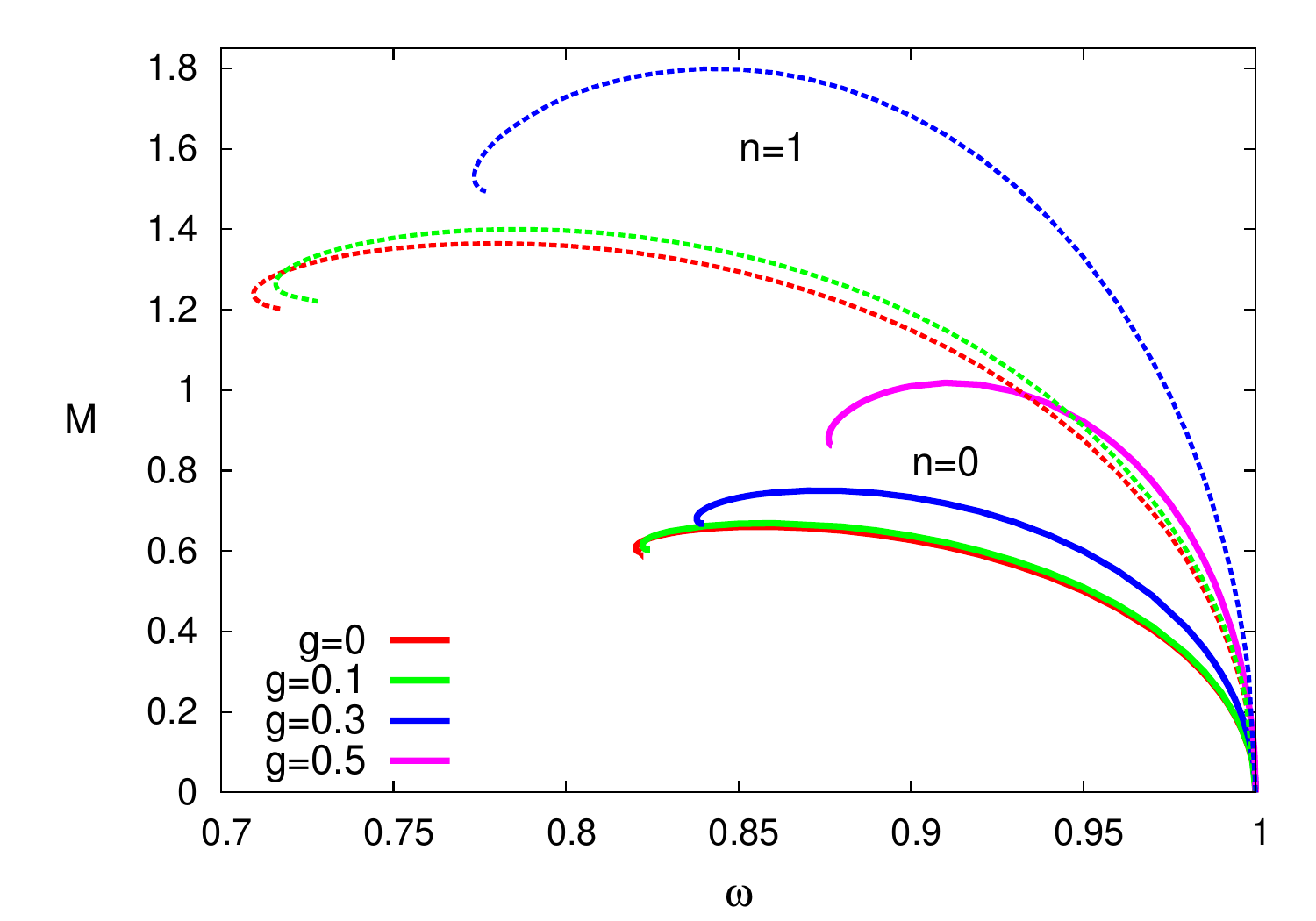}
\includegraphics[height=.33\textheight,  angle =-90]{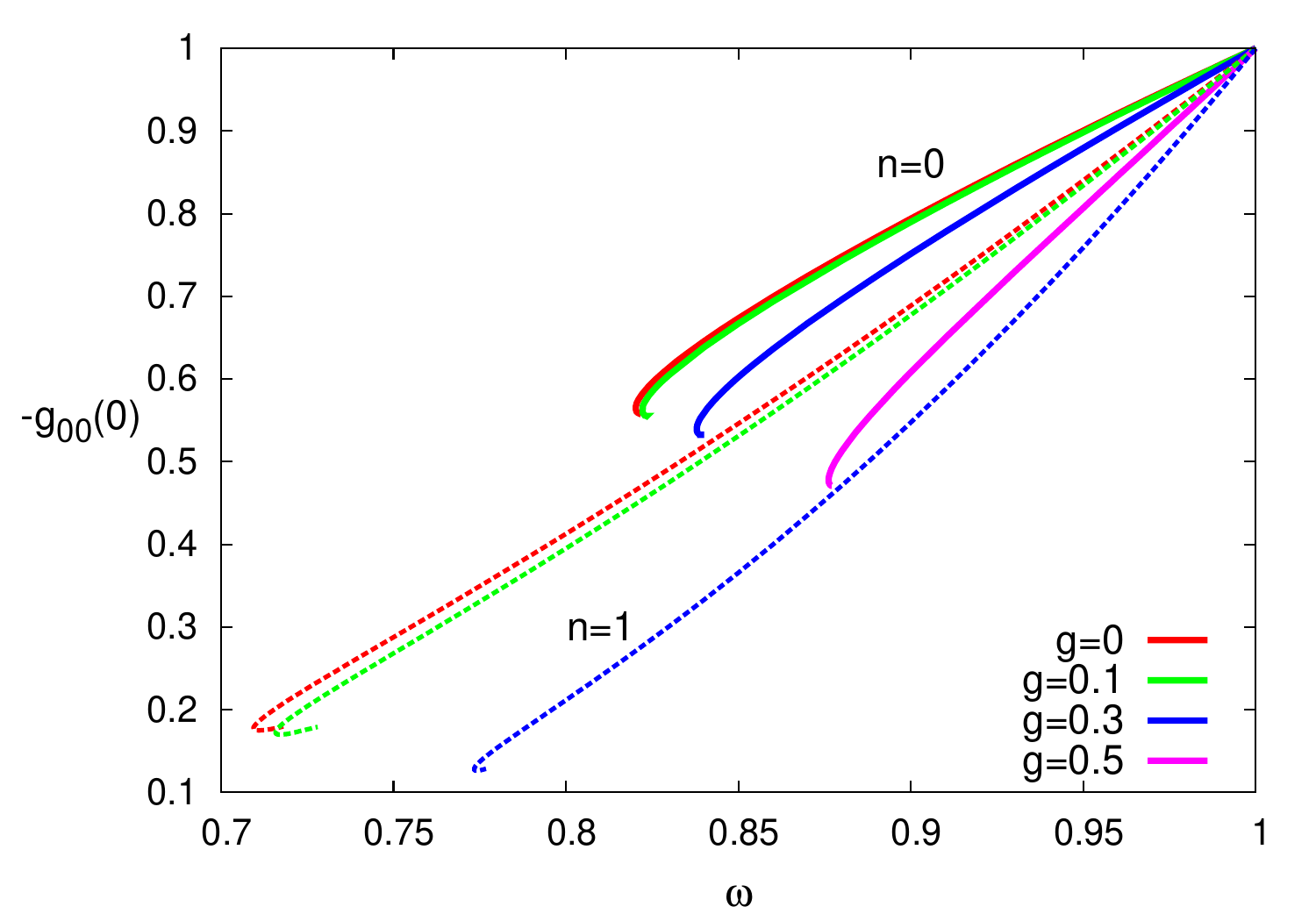}
\includegraphics[height=.33\textheight,  angle =-90]{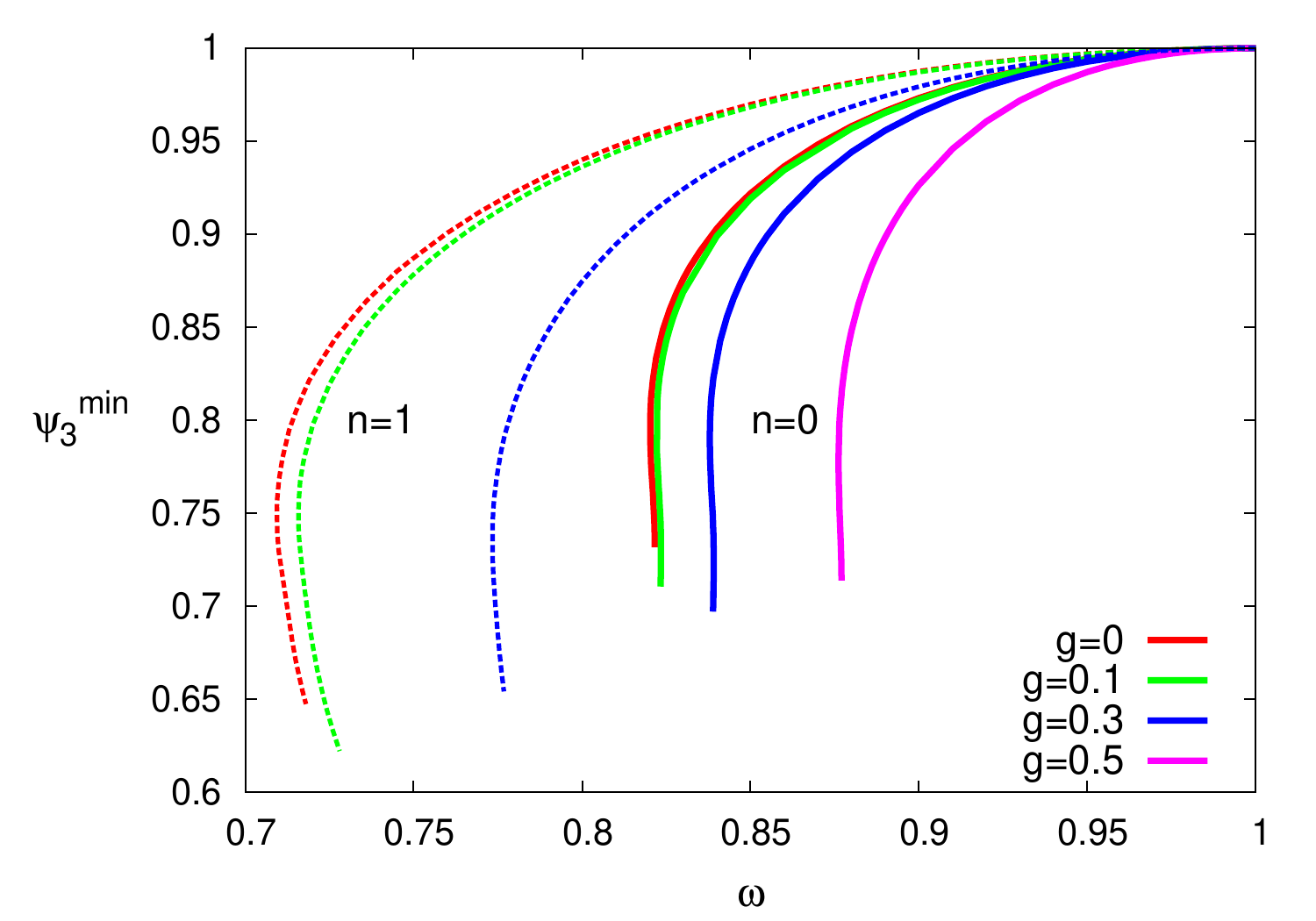}
\end{center}
\caption{\small Pion stars ($n=0$ - solid lines and $n=1$ - dotted lines): The mass $M$ (upper left plot), the value of the metric function $-g_{00}$ at the origin (upper right plot) and the minimal value of the scalar field  $\psi_3$ (bottom  plot) are displayed as functions of the frequency $\omega$ for $g=0$ and $g=0.1$\, at $\alpha=1, m=1$
}
    \lbfig{fig6}
\end{figure}
To construct the $U(1)$ gauged pion stars we start with the ungauged solutions with $g=0$ and $A_0=A_\phi=0$, described
above, and smoothly turn on the gauge interaction by increasing the value of the gauge coupling $g$, while keeping the other parameters fixed.
The domain of existence of gauged pion stars is scanned by varying the frequency $\omega$, and the effective gravitational coupling $\alpha$.
The basic properties of the gravitating gauged pion stars  can be summarized as follows:

The electromagnetic interaction shifts the fine force balance.
For any non-zero value of the gauge coupling the branches of pion star solutions cannot be extended to a limiting rescaled solution at $\alpha\to 0$, see Figs.~\ref{fig3}, \ref{fig5}.
The rescaled mass of both spherically symmetric and axially symmetric pion stars diverges at some critical value of the effective gravitational coupling $\alpha_{cr}$ below which no regular pion stars are found.
For any allowed value of $\omega$ a single branch of charged gravitating pion clouds leads to a singular strongly gravitating solution.

Given a value of $\alpha$, the minimal value of the angular frequency $\omega_{cr}$ increases with $g$, see Fig.~\ref{fig6}.
For a given values of $\omega$, gauged pion stars appear to exist up to a maximal value of the gauge coupling constant only.
Physically, this behaviour is related with the electric charge repulsion which becomes stronger as $g$ increases.
Further, the frequency-mass spiral evolution for a given $\alpha$ becomes more explicit with increasing $g$, see Fig.~\ref{fig6}.

\begin{figure}[t!]
\begin{center}
\includegraphics[height=.22\textheight,  angle =0]{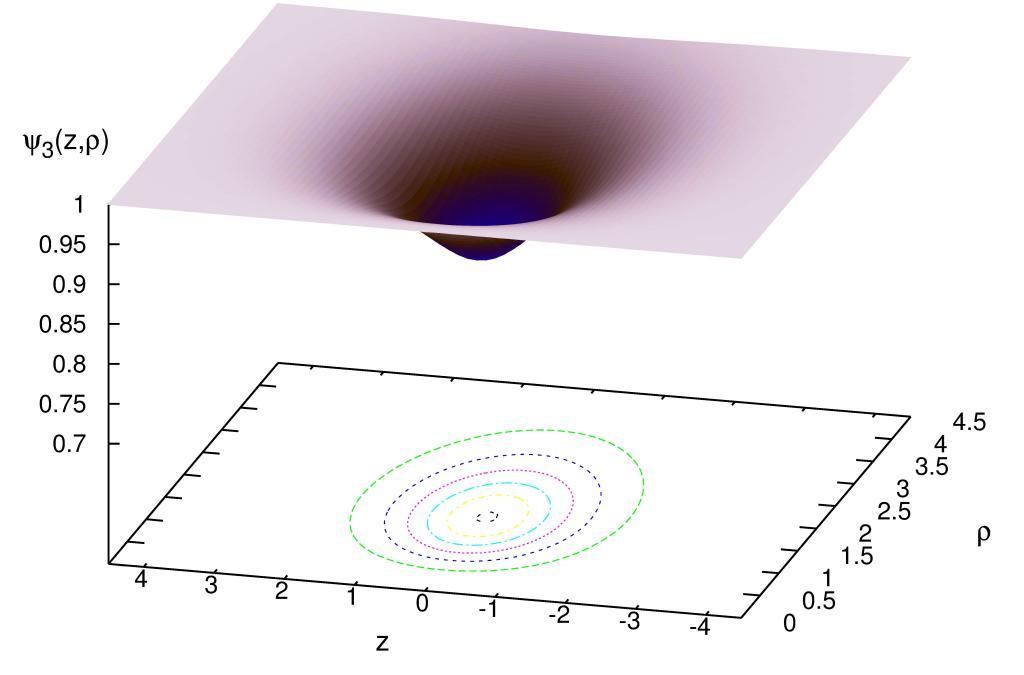}
\includegraphics[height=.22\textheight,  angle =0]{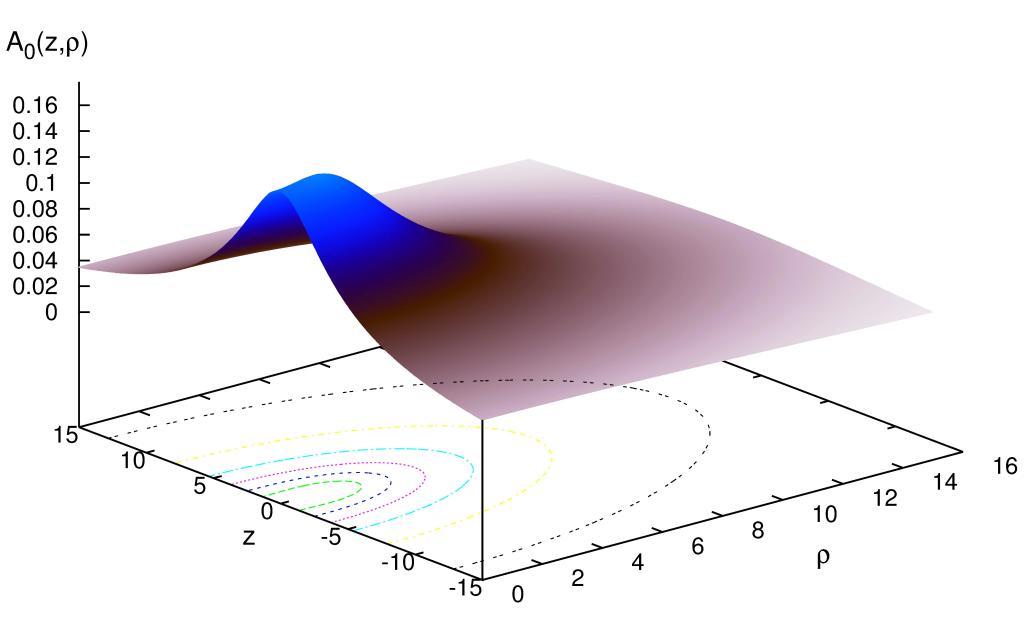}
\includegraphics[height=.22\textheight,  angle =0]{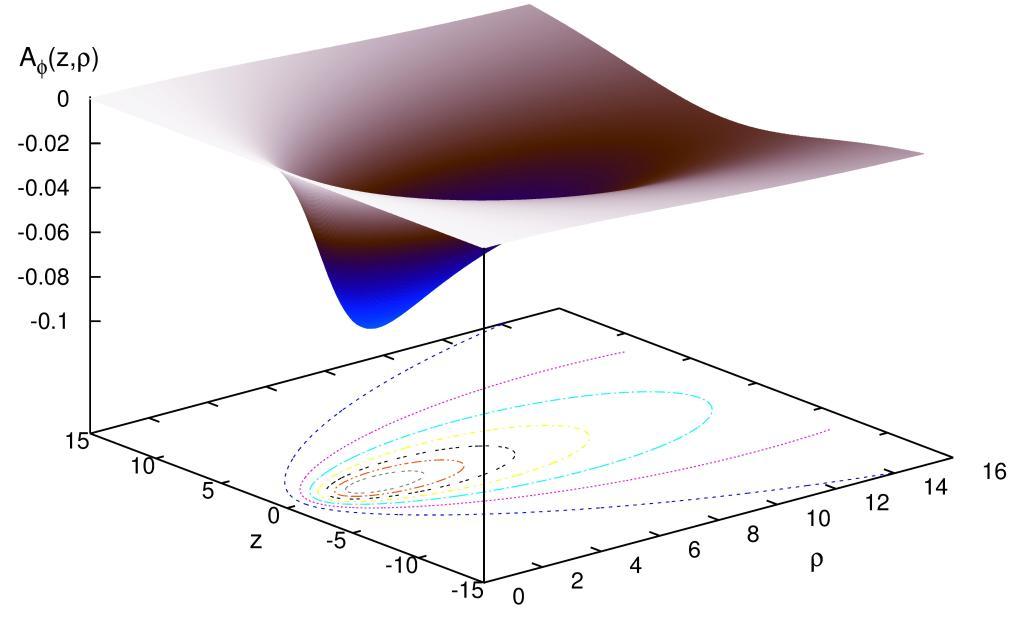}
\includegraphics[height=.22\textheight,  angle =0]{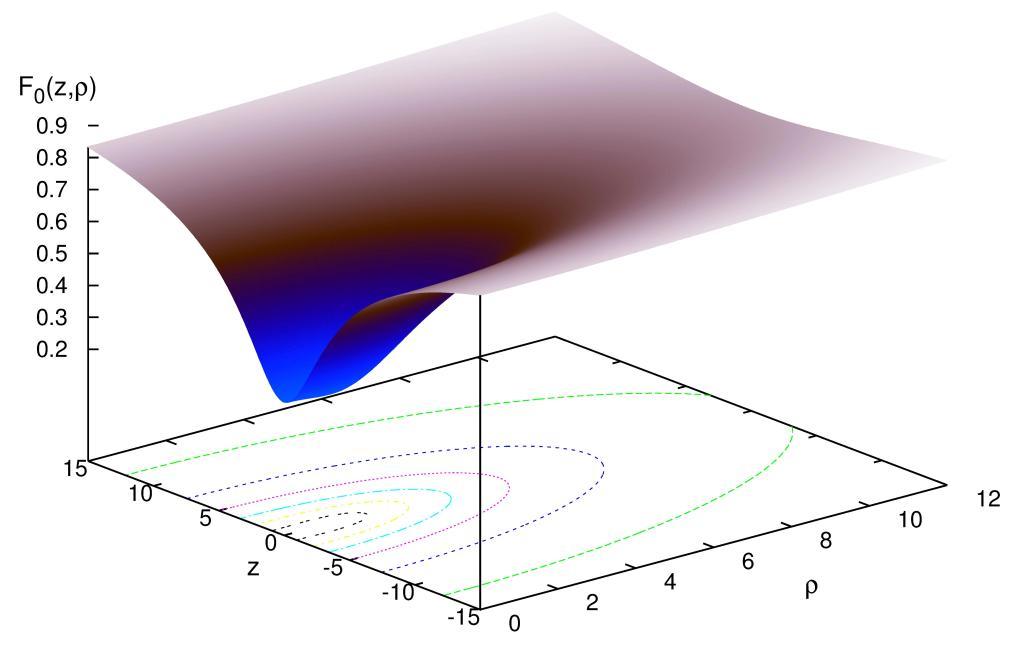}
\includegraphics[height=.22\textheight,  angle =0]{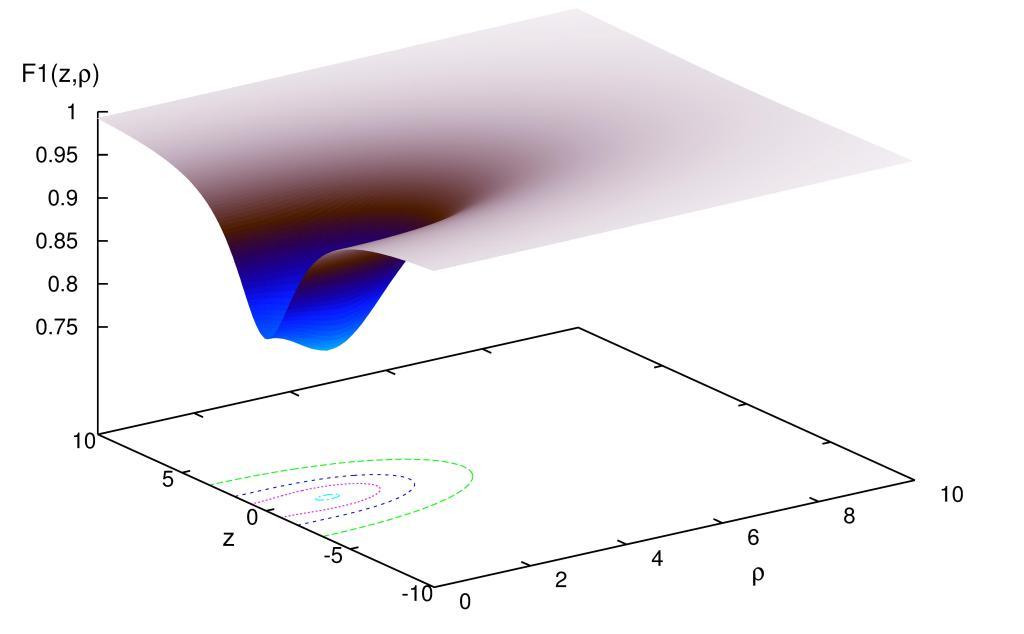}
\includegraphics[height=.22\textheight,  angle =0]{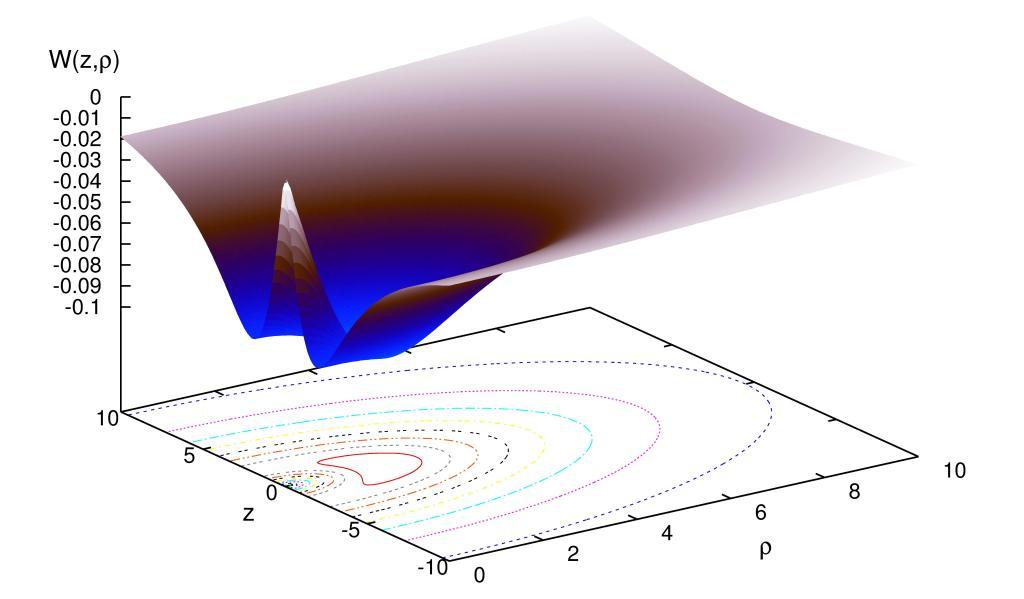}
\end{center}
\caption{\small Pion star solution ($n=1$) on the first branch: 3d plots of
the scalar field function $\psi_3$ (upper left plot),
electric potential $A_0$ (upper right plot), magnetic potential $A_\phi$ (middle left plot),  the metric
functions $F_0$ (middle right plot), $F_1$ (bottom left plot)
and $W$ (bottom right plot) versus the coordinates $\rho=r\sin\theta$
and $z=r\cos \theta$ for  $g=0.1$\, $\alpha=1, m=1$ and $\omega=0.72$
}
    \lbfig{fig7}
\end{figure}
\begin{figure}[thb!]
\begin{center}
\includegraphics[height=.20\textheight,  angle =0]{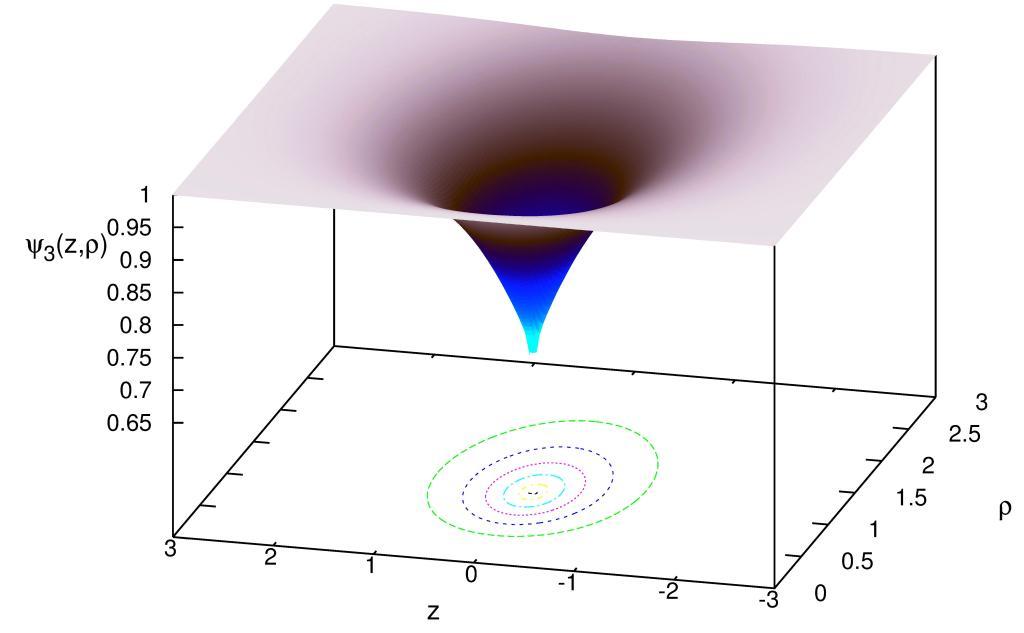}
\includegraphics[height=.20\textheight,  angle =0]{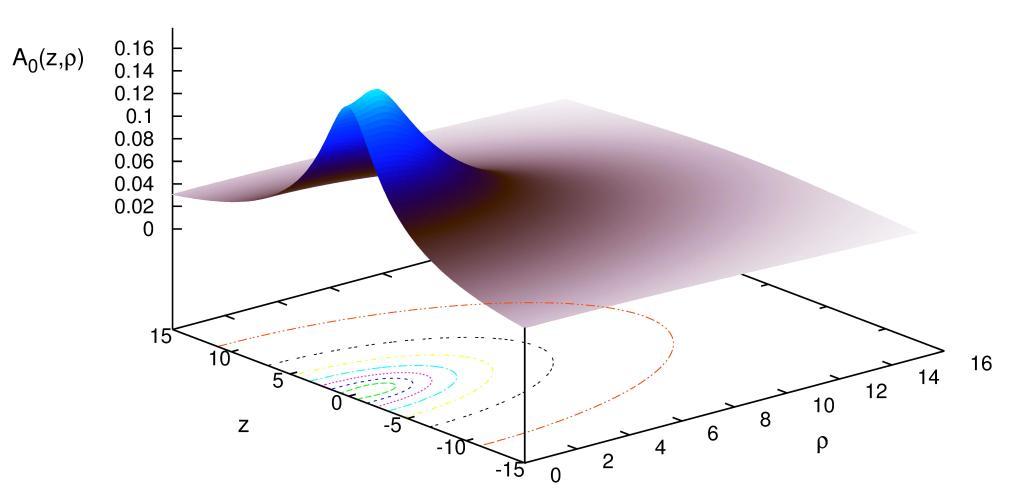}
\includegraphics[height=.21\textheight,  angle =0]{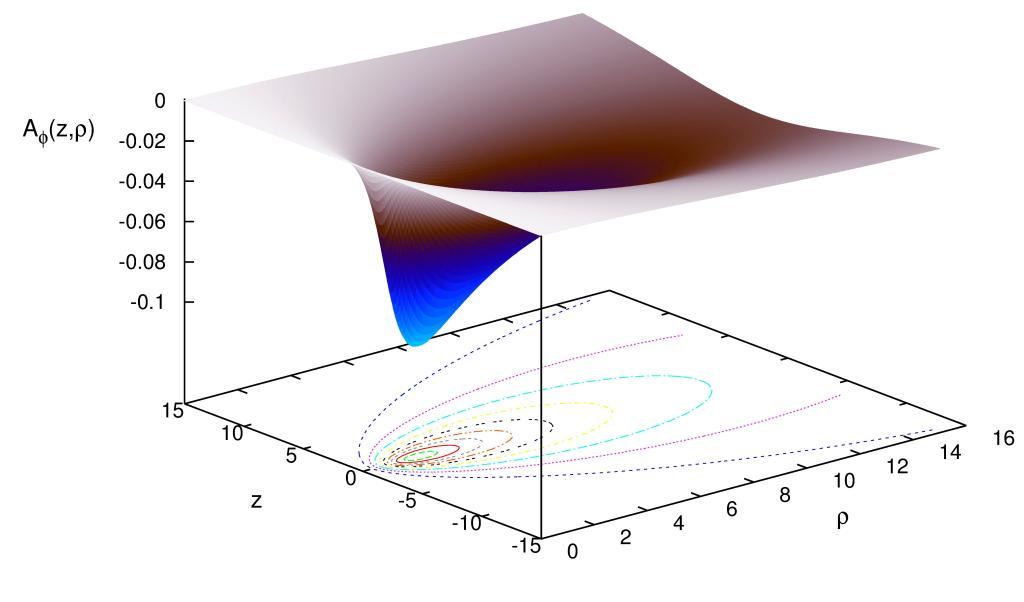}
\includegraphics[height=.21\textheight,  angle =0]{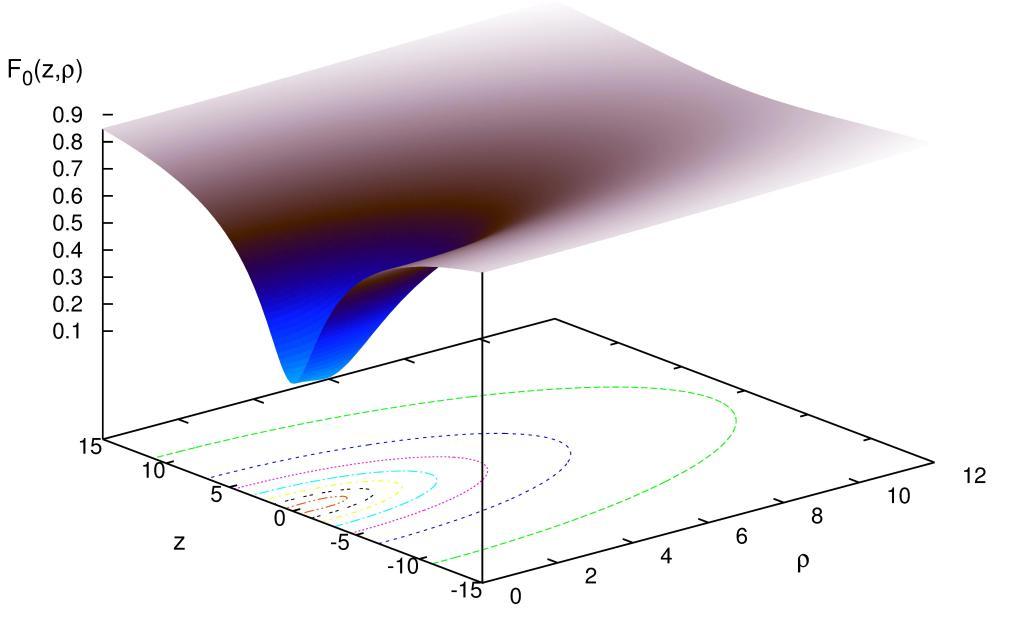}
\includegraphics[height=.21\textheight,  angle =0]{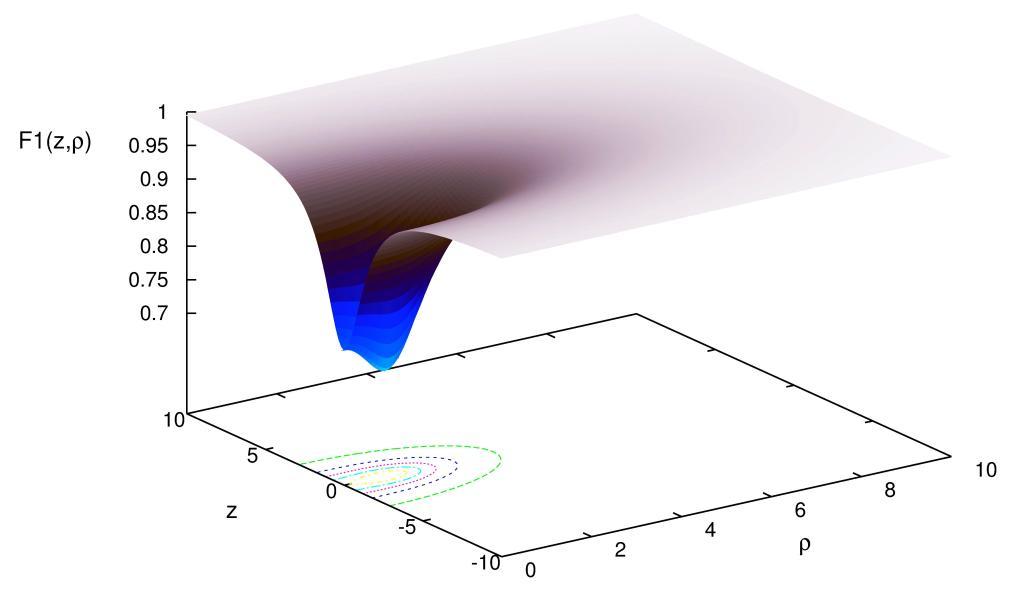}
\includegraphics[height=.21\textheight,  angle =0]{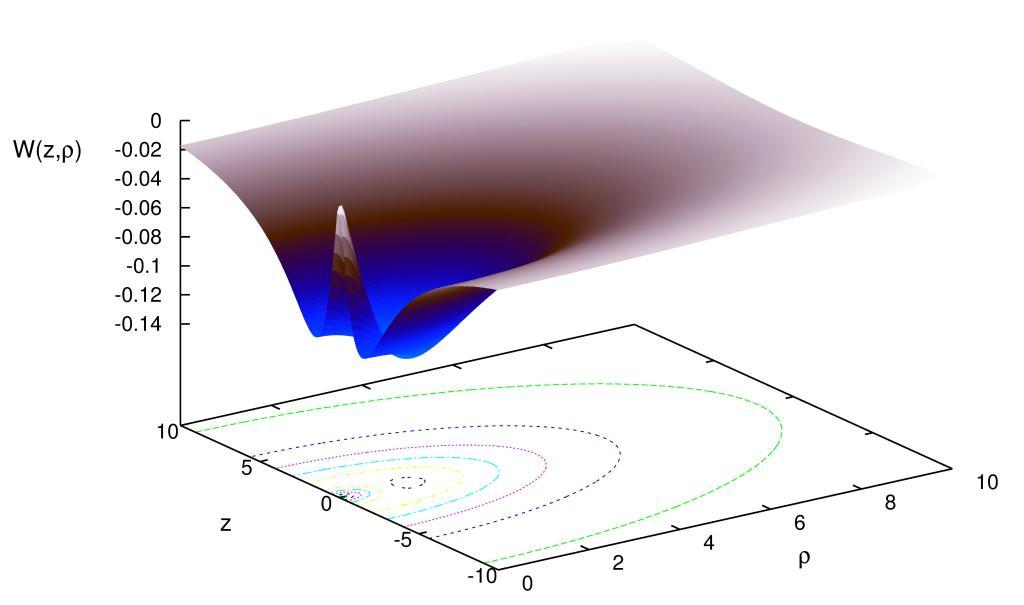}
\end{center}
\caption{\small Pion star solution ($n=1$) on the second branch: 3d plots of the scalar field function $\psi_3$ (upper left plot), electric potential $A_0$ (upper right plot), magnetic potential $A_\phi$ (middle left plot),  and the metric functions $F_0$ (middle right plot), $F_1$ (bottom left plot) and $W$ (bottom right plot) versus the coordinates $\rho=r\sin\theta$ and $z=r\cos \theta$ for  $g=0.1$\, $\alpha=1, m=1$ and $\omega=0.72$.
}
    \lbfig{fig8}
\end{figure}

As seen in Figs.~\ref{fig4}, \ref{fig6}, for any $g$, the mass-frequency dependence looks qualitatively similar to that found in the ungauged case.
The observed trend is that the maximal value of the mass $M$ increases with $g$.
Note that a similar behaviour is found for the $(\omega, J)$-dependence.

One may expect that, similar to the case of ungauged pion stars, the solutions evolve towards a limiting singular configuration.
However, numerical construction of the secondary branches becomes a very challenging numerical task, which we do not attempt in this work.
Moreover, a different numerical approach may be necessary for the systematic study of such a limit, see e.g. \cite{Kalisch:2016fkm}.

The $n=1$ axially-symmetric pion stars are electrically charged, they are also coupled to a toroidal magnetic flux which induces a magnetic dipole moment of the configuration.
In Fig.~\ref{fig7} we display particular examples of the illustrative $n=1$ solutions on the fundamental and on the second forward branch for $g = 0.1$, $\alpha = 1$, $m = 1$ and $\omega = 0.72$.
For the sake of clarity, we have chosen to exhibit these figures in polar coordinates $\rho=r \sin \theta,~ z= r\cos \theta$.
Clearly, the size of the configuration on the second branch is decreasing, the minimum of the metric component $F_0$ in the equatorial plane becomes deeper and the minimal value of the component of the scalar field $\psi_3$ decreases.
\begin{figure}[h!]
\begin{center}
\includegraphics[height=.33\textheight,  angle =-90]{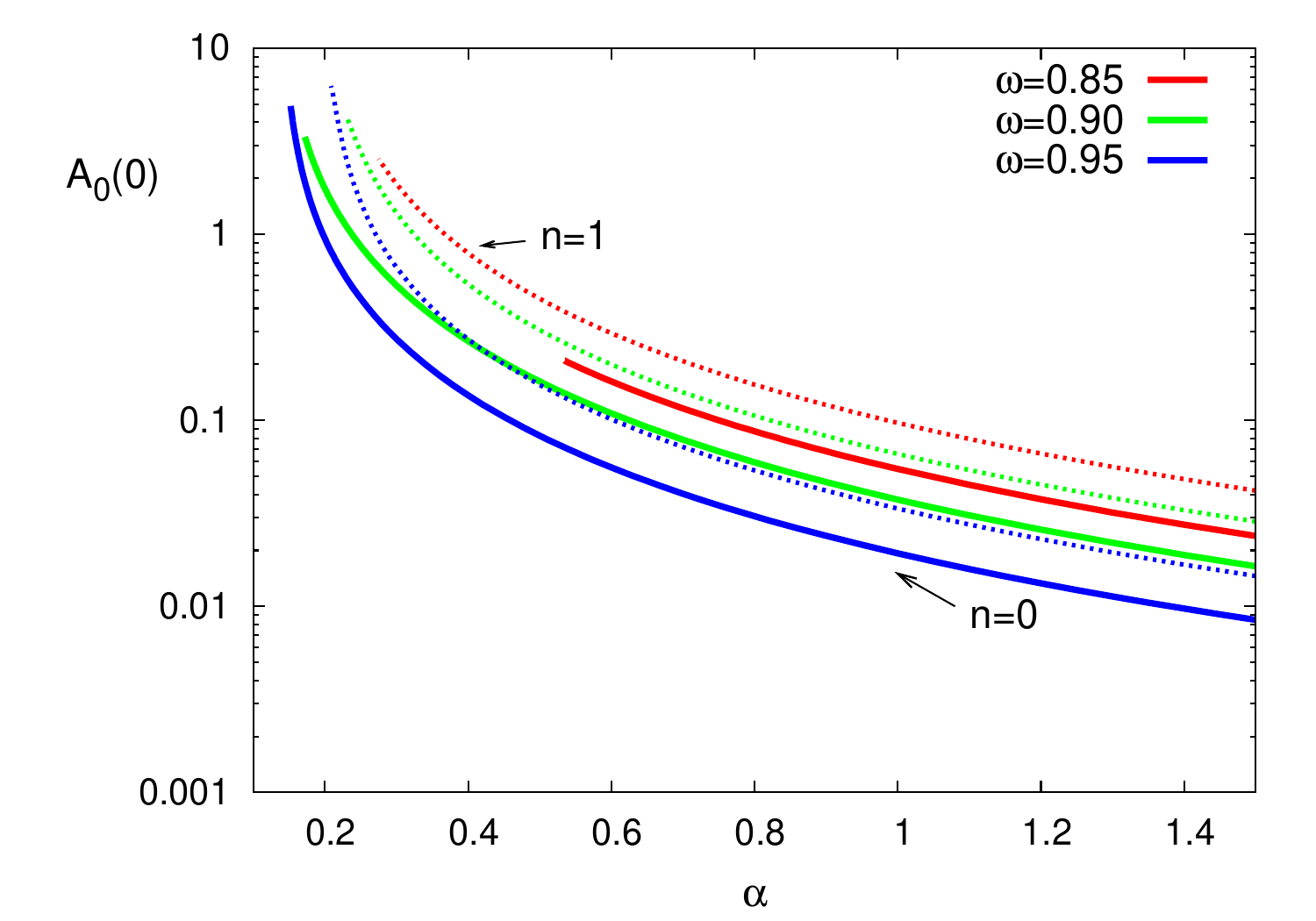}
\includegraphics[height=.33\textheight,  angle =-90]{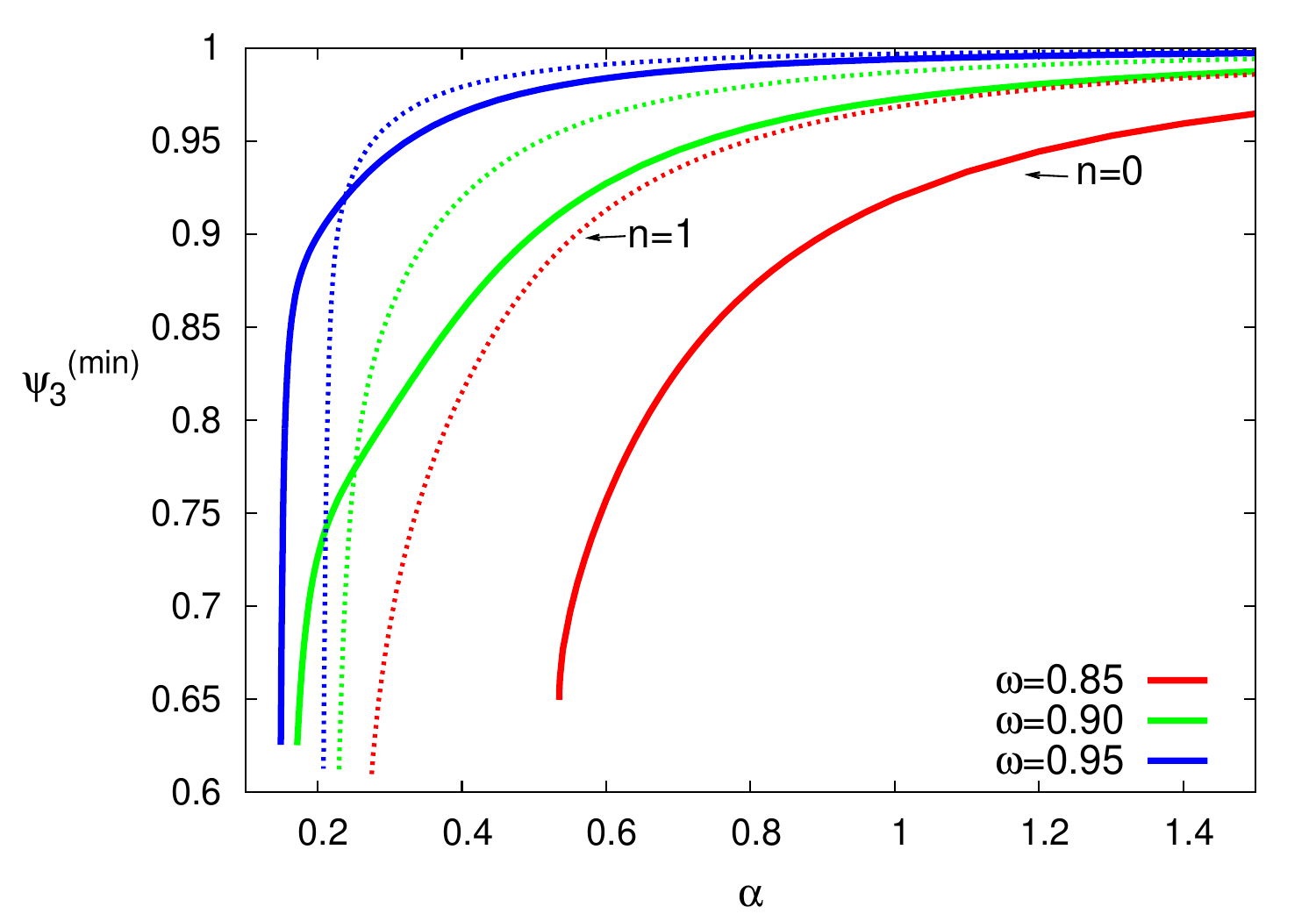}
\includegraphics[height=.33\textheight,  angle =-90]{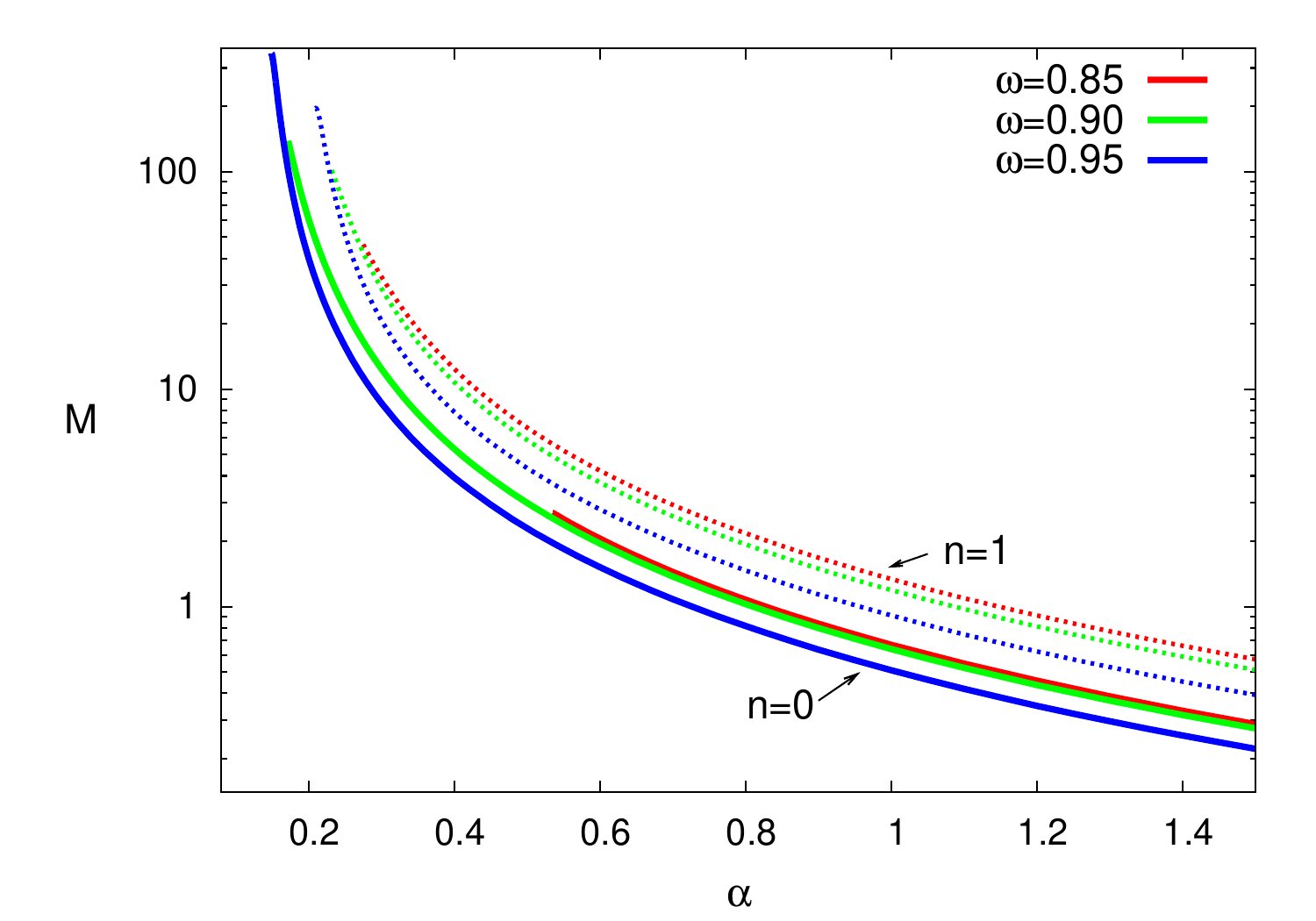}
\includegraphics[height=.33\textheight,  angle =-90]{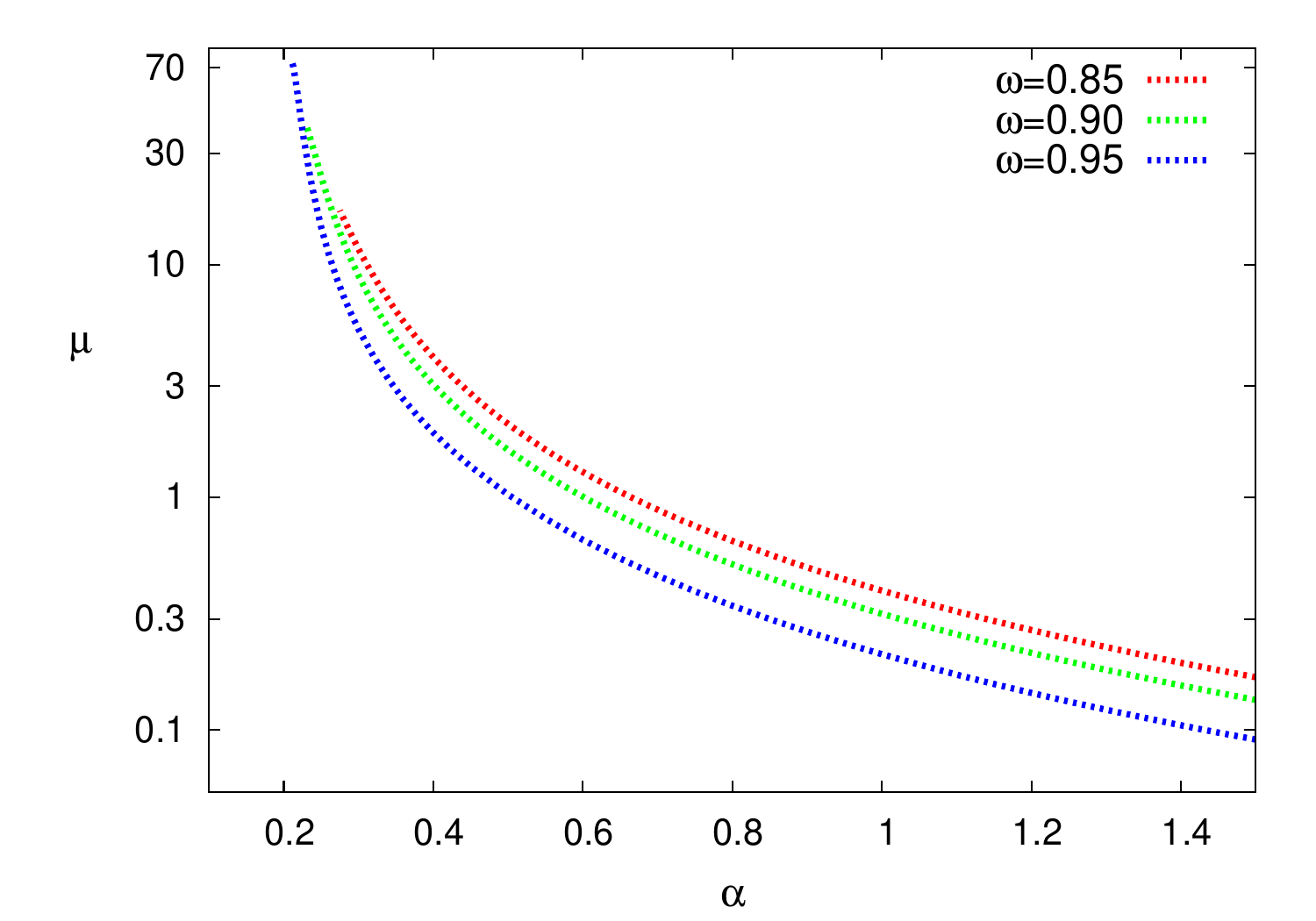}
\end{center}
\caption{\small Pion stars ($n = 0$ - solid lines and $n = 1$ - dotted lines): The gauge potential $A_0$ at the origin (upper left plot), the value of the scalar field  $\psi_3$ at the origin (upper  right plot), the ADM mass $M$ (bottom left plot) and the magnetic moment $\mu$ are displayed as functions of the effective gravitational coupling $\alpha$ for $g=0.1,\, m=1$ and several values of $\omega$.}
    \lbfig{fig9}
\end{figure}

In Fig.~\ref{fig9} we illustrate the dependence of the gauged pion stars on the effective gravitational coupling $\alpha$ for a set of values of the frequency $\omega$.
We observe, that the existence region of the charged solutions is limited by some $\alpha_{\rm min}>0$ from below, that depends on $\omega$.
When inspecting again Fig.~\ref{fig4}, where the $\omega$ dependence is illustrated for several values of $\alpha$, we see that the branches of solutions get shorter with decreasing $\alpha$.
Clearly, for a given $\omega$ there will be some $\alpha_{\rm min}>0$, for which no longer solutions exist.

Since the numerical accuracy does not allow a more detailed investigation, we now conjecture the further pattern by concluding from Fig.~\ref{fig4}.
As $\alpha$ is decreased towards this limiting value, there will be some value of $\alpha$, below which (at least) two solutions exist, since the pion stars exhibit a spiraling behavior.
(There may be even more than two solutions, but the spirals are extremely tiny here.)
Finally, at $\alpha_{\rm min}$, the two solutions on the (outer) branches of the spiral merge to the last possible solution for this value of $\omega$.
The lower critical value $\alpha_{\rm min}$ depends both on the gauge coupling $g$ and on the frequency $\omega$, and decreases as both quantities increase.

An increase of the frequency $\omega$ increases the electric charge of the configuration and the electrostatic repulsion becomes stronger, thus the attractive force must be stronger to stabilize the pion star.
When the coupling decreases 
the mass and the charge of the solutions monotonically increase, approaching  maximal values at the minimal value $\alpha_{\rm min}$, as seen in Fig.~\ref{fig9}, lower left plot.

Generally, the $n=1$ $\alpha$-branches of axially-symmetric solutions are longer than the corresponding branches of the $n=0$ pion stars.
These charged configurations are coupled to the local magnetic flux and possess a magnetic dipole moment, see Fig.~\ref{fig9}, bottom right plot.

\section{Conclusion}

In this work we have considered regular solutions of the  Einstein-Skyrme-Maxwell model, the Skyrmions of topological degree one, and non-topological configurations localized by gravity, the pion stars.
The effect of introducing the electromagnetic interaction can be summarized as follows: the increase of the gauge coupling breaks the spherical symmetry of the Skyrmion on the fundamental branch of solutions.
This effect is maximal for relatively small values of the effective gravitational coupling $\alpha$ and for large electric charge of the configuration.
By analogy with the case of ungauged Skyrmions, there are two $\alpha$-branches of solutions which bifurcate at some maximal value $\alpha_{cr}$, which increases as the gauge coupling increases.
The second backward branch extends to the limit $\alpha\to 0$.
Along this branch the electromagnetic interaction becomes less and less important and the limiting rescaled spherically symmetric Bartnik-McKinnon solution corresponds to the absence of both the Maxwell term and the quadratic kinetic term in the original Lagrangian.

We also found topologically trivial regular solutions of the Einstein-Skyrme model, whose properties are similar to those of (mini-)boson stars, or self-gravitating lumps in the non-linear $O(3)$ sigma model.
We constructed both spherically symmetric pion stars and their rotating generalizations with non-zero angular momentum and investigated their properties.

As a direction for future work, it would be interesting to study
gauged cloudy Skyrmions, multi-solitons, Skyrmed BHs with charged hair, and further configurations.

\section*{Acknowledgment}

We would like to acknowledge valuable discussions with Theodora Ioannidou, Burkhard Kleihaus and Leandro Livramento.
J.K. gratefully acknowledges support by the DFG, project Ku612/18-1.
Y.S. would like to thank the Hanse-Wissenschaftskolleg Delmenhorst for support.

 \begin{small}
 
 \end{small}

 \end{document}